\documentclass[aps,preprint,showpacs,superscriptaddress,nofootinbib]{revtex4}
 \usepackage{graphicx,amsmath,amssymb}
\usepackage[usenames]{color}
\usepackage[colorlinks=true, urlcolor=navyblue, linkcolor=navyblue, citecolor=navyblue]{hyperref}
\usepackage{relsize}
\definecolor{navyblue}{rgb}{0,0.08,0.45}
\def\Journal#1#2#3#4{{#1} {{\bf #2},} {#4} {(#3)}}

\def\PRD{{Phys. Rev.} D}
\def\PRC{{Phys. Rev.} C}

\def\ZPC{{Z. Phys. C}}

\def\MPLA{{Mod. Phys. Lett.} A}

\newcommand{\mbf}[1]{\mathbf{#1}}
\newcommand{\half}{{\frac{1}{2}}}

\def\babar{\mbox{\slshape B\kern-0.1em{\smaller A}\kern-0.1em
    B\kern-0.1em{\smaller A\kern-0.2em R}}}

\begin{document}

\preprint{SLAC-PUB-14462}

\title{Meson Transition Form Factors in Light-Front Holographic QCD}

\author{Stanley J. Brodsky}
\affiliation{SLAC National Accelerator Laboratory, Stanford University, California 94309, USA}

\author{Fu-Guang Cao}
\affiliation{Institute of Fundamental Sciences, Massey University, Private Bag 11 222, \\ Palmerston North, New Zealand}

\author{Guy F. de T{\'e}ramond}
\affiliation{Universidad de Costa Rica, San Jos\'e, Costa Rica}

\date{\today}

\begin{abstract}
We study the photon-to-meson transition form factors (TFFs) $F_{M \gamma}(Q^2)$ for  $\gamma \gamma^* \to M$ using light-front holographic methods. 
The Chern-Simons action, which is a natural form in five-dimensional anti-de Sitter (AdS) space, is required to
describe the anomalous coupling of mesons to photons using holographic methods and leads directly to
 an  expression for the photon-to-pion TFF
for a class of confining models.  Remarkably, the predicted pion TFF   is identical to the leading order QCD result where the distribution amplitude has asymptotic form. 
The Chern-Simons form is  local in AdS space and is thus somewhat limited in its predictability. It  only retains the $q \bar q$ component of the pion wave function,
and further, it projects out only the asymptotic form of the meson distribution amplitude.
It is found that in order to describe simultaneously the decay process $\pi^0 \rightarrow \gamma \gamma$ and the pion TFF  at the asymptotic limit, a probability for
the $q \bar q$ component of the pion wave function $P_{q \bar q}=0.5$ is required; thus giving indication that
the contributions from higher Fock states in the pion light-front wave function need to be included in the analysis.
The probability for the Fock state containing four quarks 
$P_{q \bar q q \bar q}  \sim 10 \%$, which follows from analyzing the hadron matrix elements for a dressed  current model, 
agrees with the analysis of the pion elastic form factor using light-front holography including higher Fock components in the pion wave function.
The results for the TFFs for the $\eta$ and $\eta^\prime$ mesons are also presented.
The rapid growth of the pion TFF exhibited by the \babar\  data at high $Q^2$ is not compatible with the models discussed in this article, whereas the theoretical calculations
are in agreement with the experimental data for the $\eta$ and $\eta^\prime$ TFFs.
\end{abstract}

\pacs{11.15.Tk, 11.25.Tq, 12.38.Aw, 13.40.Gp}
\maketitle

\section{Introduction}

The anti-de Sitter/conformal field theory (AdS/CFT) correspondence between an effective gravity theory on a higher dimensional AdS space
and conformal field theories in
physical space-time~\cite{Maldacena:1997re, Gubser:1998bc, Witten:1998qj} has led to a remarkably accurate semiclassical approximation for strongly-coupled QCD, 
and it also provides physical insights into its nonperturbative dynamics. Incorporating the AdS/CFT correspondence as a useful guide, light-front holographic methods 
were originally introduced~\cite{Brodsky:2006uqa, Brodsky:2007hb} by matching the  electromagnetic (EM) current matrix elements in AdS space~\cite{Polchinski:2002jw}
to the corresponding Drell-Yan-West (DYW) expression,~\cite{Drell:1969km, West:1970av, Soper:1976jc}  using light-front  (LF) theory in physical space-time.
One obtains the  identical holographic mapping using the matrix elements of the  energy-momentum tensor~\cite{Brodsky:2008pf} by perturbing the AdS metric
\begin{equation} \label{eq:AdSz}
ds^2 = \frac{R^2}{z^2} \left(\eta_{\mu \nu} dx^\mu dx^\nu - dz^2\right),
\end{equation}
around its static solution.~\cite{Abidin:2008ku}

A precise gravity dual to QCD is not known, but color confinement can be incorporated in the gauge/gravity correspondence by modifying the AdS geometry in the  large infrared (IR)
domain  $z \sim 1/\Lambda_{\rm QCD}$, which also sets the mass scale of the strong interactions in a class of confining models.
The modified theory generates the pointlike hard behavior expected from QCD, such as constituent counting rules~\cite{Brodsky:1973kr, Lepage:1980fj, Matveev:ra} 
 from the ultraviolet (UV) conformal limit at the AdS boundary at $z \to 0$, instead of the soft behavior characteristic of extended objects.~\cite{Polchinski:2001tt}

One can also study the gauge/gravity duality starting from the light-front
Lorentz-invariant Hamiltonian equation for the relativistic bound-state system
$P_\mu P^\mu \vert  \psi(P) \rangle =  \left(P^+ P^- \! - \mbf{P}^2_\perp\right)\vert  \psi(P) \rangle=  \mathcal{M}^2 \vert  \psi(P) \rangle$, 
$P^\pm = P^0 \pm P^3$, where the light-front time evolution operator $P^-$ is determined canonically from the QCD Lagrangian.~\cite{Brodsky:1997de}
 To a first semiclassical approximation, where quantum loops
and quark masses are not included, this leads to a LF Hamiltonian equation which describes the bound-state dynamics of light hadrons  in terms of
an invariant impact variable $\zeta$~\cite{deTeramond:2008ht}
which measures the separation of the partons within the hadron at equal light-front time
$\tau = x^0 + x^3$.~\cite{Dirac:1949cp} 
This allows us to identify the holographic variable $z$ in AdS space with the impact variable $\zeta$.~\cite{Brodsky:2006uqa, Brodsky:2008pf, deTeramond:2008ht}

The pion transition form factor (TFF) between a photon and pion measured in the $e^- e^+\to e^- e^+ \pi^0$  process, with one tagged electron,
is the simplest bound-state process in QCD.
It can be predicted from first principles in the asymptotic $Q^2 \to \infty$  limit.~\cite{Lepage:1980fj}  More generally, 
the pion TFF at large $Q^2$ can be calculated at leading twist as a convolution of a perturbative hard scattering amplitude $T_H(\gamma \gamma^* \to q \bar q)$
and a gauge-invariant meson distribution amplitude (DA) which incorporates the nonperturbative dynamics of the QCD bound-state.~\cite{Lepage:1980fj}

The \babar\ Collaboration has reported measurements of the   
transition form factors from $\gamma^* \gamma \to M$ process for the $\pi^0$,~\cite{Aubert:2009mc}
$\eta$, and $\eta^\prime$~\cite{BaBar_eta, Druzhinin:2010bg} pseudoscalar mesons for a momentum   transfer  range much larger than 
previous measurements.~\cite{CELLO,CLEO}  Surprisingly, the \babar\ data for the $\pi^0$-$\gamma$ TFF
exhibit a rapid growth for $Q^2 > 15$ GeV$^2$, which is unexpected from QCD predictions. In contrast, the data for  the  $\eta$-$\gamma$ and  $\eta'$-$\gamma$
TFFs are in agreement with previous experiments and theoretical predictions.
Many theoretical studies have been devoted to explaining \babar's experimental results.
\cite{BaBar_expln_LiM09, MikhailovS09,WuH10,BaBar_Expln_BroniowshiA10,RobertsRBGT10,PhamP11,Kroll10,GorchetinGS11,ADorokhov10,SAgaevBOP11, Brodsky:2011xx, BakulevMPS11}

Motivated by the conflict of theory with experimental results we have examined in a recent paper~\cite{Brodsky:2011xx} existing models and approximations used in the computation of pseudoscalar
meson TFFs in QCD, incorporating the evolution of the pion distribution amplitude~\cite{Lepage:1980fj, Efremov:1979qk} which controls the meson TFFs at large $Q^2$. In this article we will study the  anomalous coupling of mesons to photons which follows from the Chern-Simons (CS) action present in the dual higher dimensional gravity theory,~\cite{Witten:1998qj, Hill:2006ei}  which is required to describe the meson transition form factor using holographic principles. A simple analytical form is found  which satisfies both the low-energy 
theorem for the decay  $\pi^0 \to \gamma \gamma$ and the QCD predictions at  large $Q^2$, thus allowing us to encompass the perturbative and nonperturbative spacelike regimes in a simple model.
We choose the soft-wall approach to modify the infrared AdS geometry to include confinement, but the general results are not expected to be sensitive to the specific model chosen to deform
AdS space in the
IR since the Chern-Simons action is a topological invariant.

After a brief review of EM meson form factors in the framework of light-front holographic QCD in Sec. \ref{sec:MFF}, we discuss the Chern-Simons structure of the meson transition form factor in AdS space in Sec. \ref{sec:CSStructure}.
The pion transition form factors calculated with the free and dressed currents are presented in Sec \ref{TFFSWM}.
The higher Fock state contributions to the pion transition form factor are studied in Sec. \ref{sec:HigherTwist} for a dressed EM current model.
The results for the $\eta$ and $\eta^\prime$ transition form factors are given in Sec. \ref{sec:eta}.
Some conclusions are presented in Sec. \ref{sec:Conclusions}.
Different forms of the pion light-front wave functions (LFWFs) from holographic mappings  are discussed in the Appendix.
% \ref{sec:LFWFs}.

\section{Meson Electromagnetic Form Factor \label{sec:MFF}}

In the higher dimensional gravity theory, the hadronic transition matrix element  corresponds to
the  coupling of an external electromagnetic field $A^M(x,z)$  for a photon propagating in AdS space with the extended field $\Phi_P(x,z)$ describing a meson
in AdS~\cite{Polchinski:2002jw} and is given by
 \begin{multline} \label{MFF}
 \int d^4x \, \int dz  \sqrt{g} \, A^M(x,z)
 \Phi^*_{P'}(x,z) \overleftrightarrow\partial_M \Phi_P(x,z)
 \\ \sim
 (2 \pi)^4 \delta^{(4)} \left( P'  \! - P - q\right) \epsilon_\mu  (P + P')^\mu F_M(q^2) ,
 \end{multline}
 where the coordinates of AdS$_5$ are the Minkowski coordinates $x^\mu$ and $z$ labeled $x^M = (x^\mu, z)$,
 with $M = 1, \cdots, 5$,  and $g$ is the determinant of the metric tensor. 
The pion has initial and final four momenta $P$ and $P'$, respectively, and $q$ is the four-momentum transferred to the pion by the photon with polarization $\epsilon_\mu$.
The expression on the right-hand side
of (\ref{MFF}) represents the spacelike QCD electromagnetic transition amplitude in physical space-time
$\langle P' \vert J^\mu(0) \vert P \rangle = \left(P + P' \right)^\mu F_M(q^2)$.
It is the EM matrix element of the quark current  $J^\mu = e_q \bar q \gamma^\mu q$, and represents a local coupling to pointlike constituents. Although the expressions for the transition amplitudes look very different, one can show  that a precise mapping of the matrix elements  can be carried out at fixed light-front time.~\cite{Brodsky:2006uqa, Brodsky:2007hb}

The form factor is computed in the light front from the matrix elements of the plus-component of the current $J^+$, in order to avoid coupling to Fock states with different numbers of constituents.
Expanding the  initial and final mesons states $\vert \psi_M(P^+ \! , \mbf{P}_\perp)\rangle$ in terms of Fock components, $\vert \psi_M \rangle = \sum_n \psi_{n/M} \vert n \rangle$, we obtain
 DYW expression~\cite{Drell:1969km, West:1970av} upon the phase space integration over the intermediate variables in the $q^+=0$ frame:
 \begin{equation} \label{eq:DYW}
F_M(q^2) = \sum_n  \int \big[d x_i\big] \left[d^2 \mbf{k}_{\perp i}\right]
\sum_j e_j \psi^*_{n/M} (x_i, \mbf{k}'_{\perp i},\lambda_i)
\psi_{n/M} (x_i, \mbf{k}_{\perp i},\lambda_i),
\end{equation}
where the variables of the light-cone Fock components in the
final-state are given by $\mbf{k}'_{\perp i} = \mbf{k}_{\perp i}
+ (1 - x_i)\, \mbf{q}_\perp $ for a struck  constituent quark and
$\mbf{k}'_{\perp i} = \mbf{k}_{\perp i} - x_i \, \mbf{q}_\perp$ for each
spectator. The formula is exact if the sum is over all Fock states $n$.
The $n$-parton Fock components $\psi_{n/M} (x_i, \mbf{k}_{\perp i},\lambda_i)$ 
are independent of $P^+$ and $\mbf{P}_\perp$ and
depend only on the relative partonic coordinates:
the momentum fraction
 $x_i = k^+_i/P^+$, the transverse momentum  ${\mathbf{k}_{\perp i}}$ and spin
 component $\lambda_i^z$. Momentum conservation requires
 $\sum_{i=1}^n x_i = 1$ and
 $\sum_{i=1}^n \mathbf{k}_{\perp i}=0$.
The light-front wave functions $\psi_n$ provide a
frame-independent  representation of a hadron which relates its quark
and gluon degrees of freedom to their asymptotic hadronic state.
The form factor can also be conveniently written in impact space
as a sum of overlap of LFWFs of the $j = 1,2, \cdots, n-1$ spectator constituents~\cite{Soper:1976jc} 
\begin{equation} \label{eq:FFb}
F_M(q^2) =  \sum_n  \prod_{j=1}^{n-1}\int d x_j d^2 \mbf{b}_{\perp j}
\exp \! {\Bigl(i \mbf{q}_\perp \! \cdot \sum_{j=1}^{n-1} x_j \mbf{b}_{\perp j}\Bigr)}
\left\vert  \psi_{n/M}(x_j, \mbf{b}_{\perp j})\right\vert^2,
\end{equation}
corresponding to a change of transverse momentum $x_j \mbf{q}_\perp$ for each
of the $n-1$ spectators with  $\sum_{i = 1}^n \mbf{b}_{\perp i} = 0$.

For definiteness we shall consider 
the $\pi^+$  valence Fock state 
$\vert u \bar d\rangle$ with charges $e_u = \frac{2}{3}$ and $e_{\bar d} = \frac{1}{3}$.
For $n=2$, there are two terms which contribute to Eq. (\ref{eq:FFb}). 
Exchanging $x \leftrightarrow 1 \! - \! x$ in the second integral  we find 
\begin{equation}  \label{eq:PiFFb}
 F_{\pi^+}(q^2)  =  2 \pi \int_0^1 \! \frac{dx}{x(1-x)}  \int \zeta d \zeta \,
J_0 \! \left(\! \zeta q \sqrt{\frac{1-x}{x}}\right) 
\left\vert \psi_{u \bar d/ \pi}\!(x,\zeta)\right\vert^2,
\end{equation}
where $\zeta^2 =  x(1  -  x) \mathbf{b}_\perp^2$ and $F_{\pi^+}(q\!=\!0)=1$.

We now compare this result with the electromagnetic  form factor 
in  AdS  space-time. The incoming electromagnetic field propagates in AdS according to
$A_\mu(x^\mu ,z) = \epsilon_\mu(q) e^{-i q \cdot x} V(q^2, z)$,
where $V(q^2,z)$, the bulk-to-boundary propagator, is the solution of the AdS wave equation
given by
\begin{equation} \label{eq:V}
V(Q^2, z) = z Q K_1(z Q),
\end{equation}
with $Q^2= - q^2>0$ and boundary conditions $V(q^2 = 0, z ) = V(q^2, z = 0) = 1$.~\cite{Polchinski:2002jw}
The propagation of the pion in AdS space is described by a normalizable mode
$\Phi_P(x^\mu, z) = e^{-i P  \cdot x} \Phi(z)$ with invariant  mass $P_\mu P^\mu = \mathcal{M}_\pi^2$ and plane waves along Minkowski coordinates $x^\mu$.  
In the chiral limit for massless quarks $\mathcal{M}_\pi = 0$. 
Extracting the overall factor  $(2 \pi)^4 \delta^{(4)} \left( P'  \! - P - q\right)$ from momentum conservation at the vertex from integration over Minkowski variables in (\ref{MFF}) we find
\cite{Polchinski:2002jw} 
\begin{equation}
F(Q^2) = R^3 \int \frac{dz}{z^3} \, V(Q^2, z)  \, \Phi^2(z),
\label{eq:FFAdS}
\end{equation}
where $F(Q^2\! = 0) = 1$. 
Using the integral representation of $V(Q^2,z)$
\begin{equation} \label{eq:intJ}
V(Q^2, z) = \int_0^1 \! dx \, J_0 \! \left(\! z  Q \sqrt{\frac{1-x}{x}}\right) ,
\end{equation} we write the AdS electromagnetic form-factor as
\begin{equation} 
F(Q^2)  =    R^3 \! \int_0^1 \! dx  \! \int \frac{dz}{z^3} \, 
J_0\!\left(\!z Q\sqrt{\frac{1-x}{x}}\right) \left \vert\Phi(z) \right\vert^2 .
\label{eq:AdSFx}
\end{equation}

To compare with  the light-front QCD  form factor expression (\ref{eq:PiFFb})  we 
write the LFWF as 
\begin{equation} \label{psiphi}
\psi(x,\zeta, \varphi) = e^{i M \varphi} X(x) \frac{\phi(\zeta)}{\sqrt{2 \pi \zeta}} ,
\end{equation}
thus factoring out the angular dependence $\varphi$ in the transverse LF plane, the longitudinal $X(x)$ and
transverse mode $\phi(\zeta)$.~\footnote{The factorization of the LFWF given by (\ref{psiphi}) is a natural factorization in the light-front formalism since the
corresponding canonical generators, the longitudinal and transverse generators $P^+$ and $\mbf{P}_\perp$ and the $z$-component of the orbital angular momentum
$J^z$, are kinematical generators which commute with the LF Hamiltonian generator $P^-$.~\cite{Dirac:1949cp}} If both expressions for the form factor are identical for arbitrary values of $Q$,
we obtain $\phi(\zeta) = (\zeta/R)^{3/2} \Phi(\zeta)$ and $X(x) = \sqrt{x(1-x)}$,~\cite{Brodsky:2006uqa}
where we identify the transverse impact LF variable $\zeta$ with the holographic variable $z$,
$z \to \zeta = \sqrt{x(1-x)} \vert \mbf b_\perp \vert$.
 We choose the normalization
 $ \langle\phi\vert\phi\rangle = \int \! d \zeta \,
 \vert \langle \zeta \vert \phi\rangle\vert^2 = P_{q \bar q}$,  where $P_{q \bar q}$ is the probability of finding the $q \bar q$ component in the pion light-front wave function. The longitudinal mode is thus normalized as
$ \int_0^1 \frac{X^2(x)}{x(1-x)}=1$.~\footnote{Extension of the results to arbitrary $n$ follows from the $x$-weighted definition of the
transverse impact variable of the $n-1$ spectator system:~\cite{Brodsky:2006uqa}
$\zeta = \sqrt{\frac{x}{1-x}} ~ \Big\vert \sum_{j=1}^{n-1} x_j \mbf{b}_{\perp j} \Big\vert $,
where $x = x_n$ is the longitudinal
momentum fraction of the active quark. In general the mapping relates the AdS density $\Phi^2(z)$ to an effective LF single-particle transverse density.~\cite{Brodsky:2006uqa}}
Identical results follow from mapping the matrix elements of the energy-momentum tensor.~\cite{Brodsky:2008pf}

\subsection{Elastic form factor with a  dressed current}
\label{EFF DressedCurrent}

The results for the elastic form factor described above correspond to a ÒfreeÓ current propagating on AdS space. It is dual to the electromagnetic
pointlike current in the Drell-Yan-West
light-front formula~\cite{Drell:1969km, West:1970av} for the pion form factor.  
The DYW formula is an exact expression for the form factor. It is written as an infinite sum of an overlap of LF Fock components with an arbitrary number of constituents.
This allows one to map state-by-state to the effective gravity theory in AdS space. 
However, this mapping has the shortcoming that the multiple pole structure of the timelike form factor cannot be obtained in the timelike region unless an infinite number of Fock states is included. 
Furthermore, the moments of the form factor at  $Q^2 = 0$ diverge term-by-term; for example one obtains an infinite charge radius.~\cite{deTeramond:2011yi} 

Alternatively, one can use a truncated basis of states in the LF Fock expansion with a limited number of constituents, and the nonperturbative pole structure can be generated with
a  dressed EM current as in the Heisenberg picture, {\it i.e.},  the EM current becomes modified as it propagates in 
an IR deformed AdS space to simulate confinement.
The dressed current is dual to a hadronic EM current which includes any number of virtual $q \bar q$ components.  

Conformal invariance can be broken analytically by the introduction of a confining dilaton profile $\varphi(z)$ in the action,  $S = \int d^4 x \int dz \sqrt{g} e^{\varphi(z)} \mathcal{L},$
thus retaining conformal AdS metrics as well as  introducing a smooth IR cutoff. 
It is convenient to scale away the dilaton factor in the action by a field redefinition.~\cite{Afonin:2010hn, Lyubovitskij:2011bw}   For example, for a scalar field  we shift $\Phi \to e^{- \varphi/2} \Phi$,
and the bilinear component in the action is transformed into the equivalent problem of a free kinetic part plus an effective potential $V(\Phi, \varphi)$. 

A particularly interesting case is a dilaton profile $\exp\left(\pm \kappa^2 z^2\right)$ of either sign, since it leads to linear Regge trajectories consistent with the light-quark hadron spectroscopy.~\cite{Karch:2006pv} 
It avoids the ambiguities in the choice of boundary conditions at the infrared wall.
In  this case the
effective potential takes the form of a harmonic oscillator confining potential $\kappa^4 z^2$,
and the normalizable solution for a meson of a given twist $\tau$, corresponding to
 the lowest radial $n = 0$ node, is given by
\begin{equation}  \label{eq:Phitau}
\Phi^\tau(z) =   \sqrt{\frac{2 P_{\tau}}{\Gamma(\tau \! - \! 1)} } \, \kappa^{\tau -1} z ^{\tau} e^{- \kappa^2 z^2/2},
\end{equation} 
with normalization
\begin{equation} \label{eq:PhitauNorm}
\langle\Phi^\tau\vert\Phi^\tau\rangle = \int \frac{dz}{z^3} \, e^{- \kappa^2 z^2} \Phi^\tau(z)^2  = P_\tau,
\end{equation}
where $P_\tau$ is the probability for the twist $\tau$ mode (\ref{eq:Phitau}).
This agrees with the fact that the field $\Phi^\tau$ couples to a local hadronic interpolating operator of twist $\tau$ 
defined at the asymptotic boundary of AdS space, and thus
the scaling dimension of $\Phi^\tau$ is $\tau$. 

In the case of soft-wall potential,~\cite{Karch:2006pv}
the EM bulk-to-boundary propagator is~\cite{Brodsky:2007hb, Grigoryan:2007my}
\begin{equation} \label{eq:Vkappa}
V(Q^2,z) = \Gamma\left(1 + \frac{Q^2}{4 \kappa^2}\right) U\left(\frac{Q^2}{4 \kappa^2}, 0, \kappa^2 z^2\right),
\end{equation}
where $U(a,b,c)$ is the Tricomi confluent hypergeometric function.  The modified current $V(Q^2,z)$, Eq. (\ref{eq:Vkappa}), 
has the same boundary conditions as the free current (\ref{eq:V}),
and reduces to (\ref{eq:V}) in the  limit $Q^2 \to \infty$.
Eq.~(\ref{eq:Vkappa}) can be conveniently written in terms of the integral representation~\cite{Grigoryan:2007my}
\begin{equation}  \label{eq:Vx}
V(Q^2,z) = \kappa^2 z^2 \int_0^1 \! \frac{dx}{(1-x)^2} \, x^{\frac{Q^2}{4 \kappa^2}} 
e^{-\kappa^2 z^2 x/(1-x)}.
\end{equation} 

Hadronic form factors  for the harmonic potential $\kappa^2 z^2$  have a simple analytical form.~\cite{Brodsky:2007hb} 
Substituting in (\ref{eq:FFAdS}) the expression for a  hadronic state (\ref{eq:Phitau}) with twist $\tau = N + L$  ($N$ is the number of components)
and the bulk-to-boundary propagator (\ref{eq:Vx}) we find that the corresponding
elastic form factor for a twist $\tau$ Fock component  $F_\tau(Q^2)$
($Q^2 = - q^2 > 0$)
\begin{equation} \label{F}   
 F_\tau(Q^2) =  \frac{P_\tau}{{\Big(1 + \frac{Q^2}{M^2_\rho} \Big) }
 \Big(1 + \frac{Q^2}{M^2_{\rho'}}  \Big)  \cdots 
       \Big(1  + \frac{Q^2}{M^2_{\rho^{\tau-2}}} \Big)} ,
\end{equation}
which is  expressed as a $\tau - 1$ product of poles along the vector meson Regge radial trajectory.
For a pion, for example, the lowest Fock state -- the valence state -- is a twist-2 state, and thus the form factor is the well known monopole form.~\cite{Brodsky:2007hb}
The remarkable analytical form of (\ref{F}),
expressed in terms of the $\rho$ vector meson mass and its radial excitations, incorporates the correct scaling behavior from the constituent's hard scattering with the photon
and the mass gap from confinement.  It is also apparent from (\ref{F}) that the higher-twist components in the Fock expansion are relevant for the computation
of hadronic form factors, particularly for the timelike region which is particularly sensitive to the detailed structure of the amplitudes.~\cite{deTeramond:2010ez}
For a confined EM current in AdS a precise mapping can also be carried out to the DYW expression for the form factor. In this case we find an effective LFWF,
which corresponds to a superposition of an infinite number of Fock states. This is discussed in the Appendix for the soft-wall model.

\section{The Chern-Simons Structure of the Meson Transition Form Factor in AdS Space \label{sec:CSStructure}}

To describe the pion transition form factor within the framework of holographic QCD we need to explore the mathematical structure of higher-dimensional forms 
in the five-dimensional action, since the amplitude (\ref{MFF}) can only account for the elastic form factor $F_M(Q^2)$.
For example, in the five-dimensional compactification of type II B supergravity~\cite{Pernici:1985ju, Gunaydin:1985cu}
there is a Chern-Simons term in the action in addition to the usual Yang-Mills term $F^2$.~\cite{Witten:1998qj}
In the case of the $U(1)$ gauge theory the CS action is of the form $\epsilon^{L M N P Q} A_L \partial_M A_N \partial_P A_Q$
in the five-dimensional Lagrangian.~\cite{Hill:2006ei} 
The CS action is not gauge-invariant: under a gauge transformation it changes by a total derivative which gives a surface term.   

The Chern-Simons form is the product of three fields at the same point in five-dimensional space corresponding to a local interaction. 
Indeed the five-dimensional CS action is responsible for the anomalous coupling of mesons to photons and has been used to describe, 
for example, the $\omega \to \pi \gamma$~\cite{Pomarol:2008aa} decay as well as the 
$\gamma  \gamma^* \to \pi^0$~\cite{Grigoryan:2008up}
 and  $\gamma^* \rho^0 \to \pi^0$~\cite{Zuo:2009hz} processes.~\footnote{The anomalous EM couplings to mesons in the Sakai and Sugimoto model
 is described in Ref. \cite{Sakai:2005yt}.}

The hadronic matrix element for the anomalous electromagnetic coupling to mesons in the higher gravity theory is
given by the five-dimensional CS  amplitude
\begin{multline} \label{eq:TFFAdS1}
\int d^4 x \int dz \, \epsilon^{L M N P Q} A_L \partial_M A_N \partial_P A_Q  \\ \sim
(2 \pi)^4 \delta^{(4)} \left(P - q - k\right) F_{\pi \gamma}(q^2) \epsilon^{\mu \nu \rho \sigma} \epsilon_\mu(q) P_\nu \epsilon_\rho(k) q_\sigma,
\end{multline}
which includes the pion field as well as the external photon fields by identifying the fifth component of $A$ with the meson mode in AdS space.~\cite{Hill:2004uc}
In the right-hand side of (\ref{eq:TFFAdS1})  $q$ and $k$ are the momenta of the virtual and on-shell incoming photons respectively 
with  corresponding polarization vectors  $\epsilon_\mu(q)$ 
and $\epsilon_\mu(k)$ for  the amplitude   $\gamma \gamma^* \to \pi^0$.
The momentum of the outgoing  pion is $P$.

The pion transition form factor $F_{\pi \gamma}(Q^2)$  can be computed from first principles in QCD. To leading 
leading order in $\alpha_s(Q^2)$ and leading twist the result is~\cite{Lepage:1980fj} ($Q^2 = - q^2 >0$)
\begin{equation}
Q^2 F_{\pi \gamma}(Q^2)=\frac{4}{\sqrt{3}} \int_0^1  {\rm d} x \frac{\phi(x,{\bar x} Q)}{\bar x}
\left[ 1+ O \left(\alpha_s,\frac{m^2}{Q^2} \right) \right],
\label{eq:TFLB1}
\end{equation}
where $x$ is the longitudinal momentum fraction of the quark struck by the virtual photon in the hard scattering process
and ${\bar x}=1-x$ is the longitudinal momentum fraction of the spectator quark.  
The pion distribution amplitude $\phi(x,Q)$ in the light-front formalism~\cite{Lepage:1980fj} is the integral of the 
valence $q \bar q$ LFWF in light-cone gauge $A^+=0$
\begin{equation}
\phi(x,Q)=\int_0^{Q^2}\frac{ d^2 \mbf{k}_\perp}{16 \pi^3} \psi_{q \bar q/ \pi}(x, \mbf{k}_\perp),
\label{eq:DALC}
\end{equation}
and has the asymptotic form~\cite{Lepage:1980fj}  $\phi(x, Q \to \infty) = \sqrt{3} f_\pi x (1-x)$; thus the  leading order  QCD result  for the TFF at the asymptotic limit
is obtained,~\cite{Lepage:1980fj}
\begin{equation} \label{TFFasy}
Q^2 F_{\pi \gamma}(Q^2 \rightarrow \infty)=2 f_\pi.
\end{equation}

We now compare the QCD expression on the right-hand side of  (\ref{eq:TFFAdS1}) with the AdS transition amplitude on the left-hand side
As for the elastic form factor discussed in Sec. \ref{sec:MFF}, the incoming off-shell photon is represented by the propagation of the non-normalizable electromagnetic solution
 in AdS space,
$A_\mu(x^\mu ,z) = \epsilon_\mu(q) e^{-i q \cdot x} V(q^2, z)$,
where $V(q^2,z)$ is the bulk-to-boundary propagator with boundary conditions $V(q^2 = 0, z ) = V(q^2, z = 0) = 1$.~\cite{Polchinski:2002jw}
Since the incoming photon with momentum $k$ is on its mass shell, $k^2 = 0$,  its wave function is $A_\mu(x^\mu, z) = \epsilon_\mu(k) e^{ - i k \cdot x}$.   
Likewise, the propagation of the pion in AdS space is described by a normalizable mode
$\Phi_{P}(x^\mu, z) = e^{-i P  \cdot x} \Phi_\pi(z)$ with invariant  mass $P_\mu P^\mu = \mathcal{M}_\pi^2 =0$   
in the chiral limit for massless quarks. 
The normalizable mode $\Phi(z)$ scales as $\Phi(z) \to z^{\tau=2}$  in the limit $z \to 0$, since the leading interpolating operator for the pion has twist-2.
 A simple dimensional analysis implies that  $A_z \sim  \Phi_\pi(z)/ z$, matching the twist scaling dimensions: two for the pion and one for the EM field. 
 Substituting in (\ref{eq:TFFAdS1}) the expression given above for the 
 pion and the EM fields  propagating in  AdS,   and extracting the overall factor  $(2 \pi)^4 \delta^{(4)} \left( P  \! - q - k\right)$
 upon integration over Minkowski variables in (\ref{eq:TFFAdS1}) we find  $(Q^2 = - q^2 > 0)$
\begin{equation}  \label{eq:TFFAdS2}
F_{\pi \gamma}(Q^2) = \frac{1}{2 \pi} \int_0^\infty   \frac{d z}{z} \,  \Phi_\pi(z)  V\!\left(Q^2, z\right) ,
\end{equation}
where the normalization is fixed by the asymptotic QCD prediction (\ref{TFFasy}). We have defined our units such that the AdS radius $R=1$. 

Since the LF mapping of (\ref{eq:TFFAdS2}) to the asymptotic QCD prediction (\ref{TFFasy}) only depends on the asymptotic behavior near the boundary of AdS space,
 the result is independent of the particular model used to 
modify the large $z$ IR region of AdS space.  At large enough $Q$, the important contribution to (\ref{TFFasy}) only comes from the region near $z \sim 1/Q$ where 
$\Phi(z) = 2 \pi f_\pi z^2 + \mathcal{O}(z^4)$.  Using the integral
\begin{equation}
\int_0^\infty dx \, x^\alpha K_1(x) = 2^{\alpha -2} \alpha \, \left[\Gamma \! \left(\frac{\alpha}{2}\right)\right]^2,
~~~ {\rm Re}(\alpha) >1,
\end{equation}
we recover the asymptotic result  (\ref{TFFasy})
\begin{equation}
Q^2 F_{\pi \gamma}(Q^2 \rightarrow \infty)=2 f_\pi + \mathcal{O}\left(\frac{1}{Q^2}\right),
\end{equation} 
with the pion decay constant $f_\pi$ (See Appendix \ref{appendix})
\begin{equation} \label{eq:fpi}
f_\pi = \frac{1}{4 \pi} \frac{\partial_z\Phi_\pi(z)}{z} \Big\vert_{z=0}.
\end{equation}
Since the pion field is identified as the fifth component of $A_M$,
the CS form $\epsilon^{L M N P Q} A_L \partial_M A_N \partial_P A_Q$ is similar in form to an axial current;  this correspondence can explain 
why the resulting pion distribution amplitude has the asymptotic form. 

In Ref. \cite{Grigoryan:2008up} the pion TFF was studied in the framework of a CS extended hard-wall AdS/QCD model with  $A_z \sim \partial_z \Phi(z)$.  The  expression for the TFF
which follows from (\ref{eq:TFFAdS1}) then vanishes at $Q^2 =0$, and has to be corrected by the introduction of a surface term at the IR wall.~\cite{Grigoryan:2008up}
However, this procedure is only possible for a model with a sharp cutoff.
The pion TFF has also been studied using the holographic approach to  QCD in Refs.~\cite{Cappiello:2010uy, Stoffers:2011xe,RodriguezGomez:2008zp}.

\section{A Simple Holographic Confining Model \label{TFFSWM}}

QCD  predictions of the TFF  correspond to the local coupling of the free electromagnetic current to the elementary constituents
in the interaction representation.~\cite{Lepage:1980fj} To compare with  QCD results, we first consider a simplified model where the non-normalizable mode $V(Q^2,z)$
for the  EM current satisfies the  ``free" AdS equation
subject to the boundary conditions $V(Q^2 =0, z) = V(Q^2, z =0) = 1$; thus the solution  $V(Q^2, z) = z Q K_1(z Q)$,
dual  to the free electromagnetic current.~\cite{Brodsky:2006uqa} To describe the normalizable mode representing the pion we take the soft-wall exponential form (\ref{eq:Phitau}).
Its LF mapping has also a convenient exponential form and has been studied considerably in the literature.~\cite{Brodsky:2011xx}   The exponential form of the LFWF
in momentum space has important support only when the virtual states are near the energy shell, and thus it  implements in a natural way the requirements of
the bound-state dynamics. From (\ref{eq:Phitau}) we have for  twist $\tau = 2$
\begin{equation}   \label{eq:Phipi}
\Phi_{q \bar q/ \pi} (z) =   \sqrt{2 P_{q \bar q}} \, \kappa \, z^2 e^{- \kappa^2 z^2/2},
\end{equation} 
with normalization
\begin{equation}
\langle\Phi_{q \bar q/ \pi} \vert \Phi_{q \bar q/ \pi} \rangle 
= \int \frac{dz}{z^3} \, e^{- \kappa^2 z^2} \Phi_{q \bar q/ \pi}^2(z)  = P_{q \bar q},
\end{equation}
where $P_{q \bar q}$ is the probability for the valence state.
From (\ref{eq:fpi}) the pion decay constant  is
\begin{equation} \label{eq:fpiPhi}
f_\pi = \sqrt{P_{q \bar q} } \, \frac{\kappa}{\sqrt{2} \pi}.
\end{equation}
 It is not possible in this model to introduce a surface term as in Ref. \cite{Grigoryan:2008up} to match the
value of the TFF at $Q^2 = 0$  derived from the decay $\pi^0 \to \gamma \gamma$. Instead, higher Fock components which modify the pion wave function at large distances
 are required 
to satisfy this low-energy constraint naturally. Since the higher-twist components  have a faster fall-off at small distances, the asymptotic results are not modified.

 Substituting  the pion wave function (\ref{eq:Phipi}) and using the integral representation for $V(Q^2, z)$
\begin{equation}
z Q K_1(z Q) = 2 Q^2 \int_0^\infty \frac{t J_0(z t)}{(t^2 + Q^2)^2} dt ,
\end{equation}
we find upon integration
\begin{equation}
F_{\pi \gamma}(Q^2) = \frac{\sqrt {2 P_{q \bar q}} ~ Q^2 }{\pi \kappa} \int_0^\infty \frac{t dt }{(t^2  + Q^2)^2} e^{- t^2 / 2 \kappa^2}.
\end{equation}
Changing variables as  $x = \frac{Q^2}{t^2 + Q^2}$ one obtains
\begin{equation}
F_{\pi \gamma}(Q^2) = \frac{P_{q \bar q}}{2 \pi^2 f_\pi} \int_0^1 dx \exp \left( - \frac{(1-x) P_{q \bar q} Q^2 }{4 \pi^2 f_\pi^2  x}\right).
\label{eq:FpiFCT2}
\end{equation}
Upon integration by parts, Eq. (\ref{eq:FpiFCT2})  can also be written as 
\begin{equation} \label{eq:TFFAdSQCD}
Q^2 F_{\pi \gamma}(Q^2) = \frac{4}{\sqrt{3}} \int_0^1 dx  \frac{\phi(x)}{1-x}  \left[1 - \exp \left( - \frac{(1-x) P_{q \bar q} Q^2 }{4 \pi^2 f_\pi^2  x}\right) \right] ,
\end{equation}
where $\phi(x) = \sqrt{3} f_\pi x(1-x)$ is the asymptotic QCD distribution amplitude with $f_\pi$ given by
(\ref{eq:fpiPhi}).

Remarkably, the pion transition form factor given by (\ref{eq:TFFAdSQCD}) 
for $P_{q \bar q} =1$
is identical to the 
results for the pion TFF obtained with the exponential light-front wave function model of 
Musatov and Radyushkin~\cite{Musatov:1997pu} consistent with the leading order  QCD result~\cite{Lepage:1980fj} for the TFF at the asymptotic limit,  
$Q^2 F_{\pi \gamma}(Q^2 \rightarrow \infty)=2 f_\pi$.~\footnote{The expression (\ref{eq:TFFAdSQCD}) is not appropriate to describe the timelike region
where the exponential factor in (\ref{eq:TFFAdSQCD}) 
grows exponentially. It is important to study the behavior of the pion TFF in other kinematical regions to describe, for example, the process $e^+ + e^- \to \gamma^* \to \pi^0 + \gamma$. 
This also would test the \babar\ anomaly.}~\footnote{A similar mapping can be done for the case when the two photons are virtual $\gamma^* \gamma^* \to \pi^0$.
In the case where at least one of the incoming photons has
large virtuality the transition form factor can be expressed analytically in a simple form. The result is 
$F_{\pi \gamma^*}(q^2, k^2) = - \frac{4}{\sqrt{3}} \int_0^1 dx  \frac{\phi(x)}{x q^2 + (1-x) k^2} $, with $\phi(x)$  the asymptotic DA. See Ref.~\cite{Grigoryan:2008up}.}
The leading-twist result  (\ref{eq:TFFAdSQCD}) does not include nonleading order  $\alpha_s$ corrections in the hard scattering amplitude nor gluon exchange in the evolution of
 the distribution amplitude, since the semiclassical correspondence implied in the gauge/gravity duality does not contain quantum effects such as particle emission and absorption. 

The transition form factor at $Q^2=0$ can be obtained from Eq.~(\ref{eq:TFFAdSQCD}),
\begin{equation}
F_{\pi \gamma}(0) = \frac{1}{2  \pi^2 f_{\pi}} P_{q \bar q}.
\label{eq:pionTFFFCT2Q0}
\end{equation}
The form factor $F_{\pi \gamma}(0)$ is related to the decay width for the $\pi^0 \rightarrow \gamma \gamma$ decay,
\begin{equation}
\Gamma_{\pi^0 \rightarrow \gamma \gamma}=\frac{\alpha^2 \pi m_\pi^3}{4} F_{\pi \gamma}^2(0),
\end{equation}
where $\alpha=1/137$.
The form factor $F_{\pi \gamma}(0)$ is also well described by the Schwinger, Adler, Bell and Jackiw anomaly~\cite{Schwinger:1951nm} which gives
\begin{equation}
F_{\pi \gamma}^{\rm SABJ}(0) = \frac{1}{4 \pi^2 f_\pi},
\label{eq:JackiwAnomaly}
\end{equation}
 in agreement within a few percent of  the observed value obtained from
the decay $\pi^0 \to \gamma \gamma$.

\begin{figure}[htbp]
\begin{center}
\includegraphics[width=9cm]{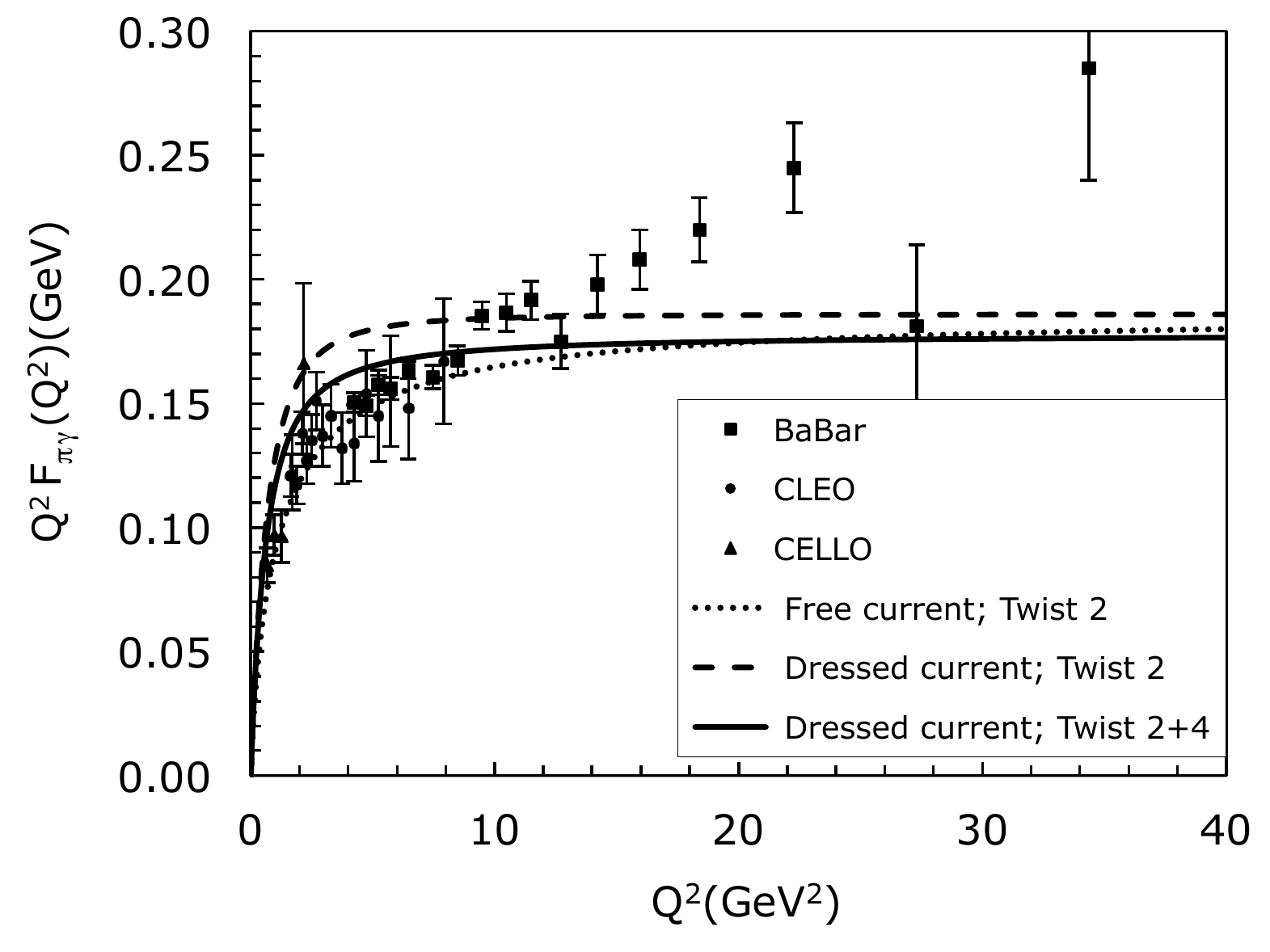}
\caption{The $\gamma \gamma^* \to \pi^0$ transition form factor shown as $Q^2 F_{\pi \gamma}(Q^2)$ as a function of $Q^2 = -q^2$.
The dotted curve is the asymptotic result predicted by the Chern-Simons form. 
The dashed and solid curves include the effects of using a confined EM current for twist-2 and twist-2 plus twist-4 respectively. The data are from \cite{Aubert:2009mc, CELLO,  CLEO}. }
\label{Q2PiTFF}
\end{center}
\end{figure}

\begin{figure}[htbp]
\begin{center}
\includegraphics[width=9cm]{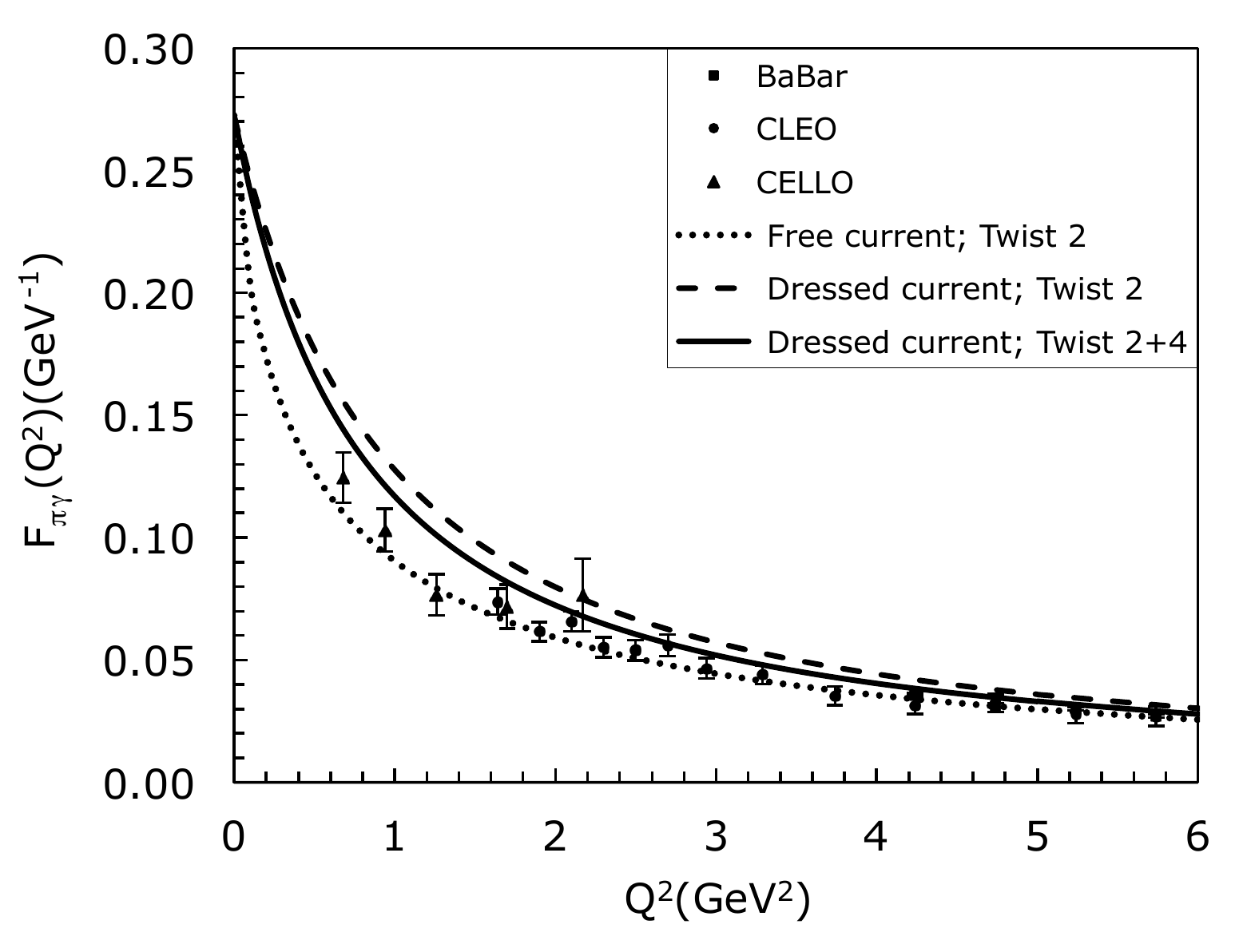}
\caption{Same as Fig.~\ref{Q2PiTFF} for $F_{\pi \gamma}(Q^2)$. }
\label{PiTFF}
\end{center}
\end{figure}

 Taking $P_{q \bar q}=0.5$ in (\ref{eq:pionTFFFCT2Q0}) one obtains a  result in agreement with (\ref{eq:JackiwAnomaly}).
 This suggests that the contribution from higher Fock states vanishes at $Q=0$ in this simple holographic confining model (see Sec. \ref{sec:HigherTwist} for further discussion).
 Thus (\ref{eq:TFFAdSQCD}) represents  a description on the pion TFF
  which encompasses the low-energy nonperturbative  and the high-energy hard domains, but includes only
  the asymptotic DA of the $q \bar q$ component of the pion wave function at all scales.
The results from  (\ref{eq:TFFAdSQCD}) are shown as dotted curves in Figs.  \ref{Q2PiTFF} and  \ref{PiTFF} for $Q^2 F_{\pi\gamma}(Q^2)$ and $F_{\pi\gamma}(Q^2)$ respectively.
The calculations agree reasonably well with the experimental data at low- and medium-$Q^2$ regions ($Q^2<10$ GeV$^2$) , but disagree with \babar's large $Q^2$ data.

\subsection{Transition form factor with the dressed current}
\label{sec:TFFDressedCurrent}

The simple valence $q \bar q$ model discussed above should be modified at small $Q^2$ by introducing the  dressed current  which corresponds effectively to a superposition of Fock states (see the Appendix).
% \ref{appendix}).
Inserting the valence pion wave function (\ref{eq:Phipi})  and the confined EM current  (\ref{eq:Vx}) in 
the amplitude (\ref{eq:TFFAdS2})
one finds 
\begin{equation}  \label{eq:DressedFpiT2}
F_{\pi \gamma}(Q^2) = \frac{P_{q \bar q} }{\pi^2 f_\pi} \int_0^1  \frac{d x }{(1+x)^2} \, x^{Q^2 P_{q \bar q}/(8 \pi^2 f_\pi^2)}.
\end{equation}
Equation~(\ref{eq:DressedFpiT2}) gives the same value for $F_{\pi \gamma}(0)$ as (\ref{eq:pionTFFFCT2Q0}) which was  obtained with the free current.
Thus the anomaly result $F_{\pi \gamma}(0)=1/(4 \pi^2 f_\pi)$ is reproduced if $P_{q \bar q}=0.5$ is also taken in (\ref{eq:DressedFpiT2}).
 Upon integration by parts, Eq.~(\ref{eq:DressedFpiT2}) can also be written as
\begin{equation}
Q^2 F_{\pi \gamma}(Q^2) = 8 f_\pi  \int_0^1  d x \frac{1- x }{(1+x)^3} \, \left( 1- x^{Q^2 P_{q \bar q}/(8 \pi^2 f_\pi^2)} \right).
\label{eq:DressedFpiT2Asy}
\end{equation}
Noticing that the second term in Eq.~(\ref{eq:DressedFpiT2Asy}) vanishes at the limit $Q^2 \rightarrow \infty$,
one recovers Brodsky-Lepage's asymptotic prediction for the pion TFF:  $Q^2 F_{\pi \gamma}(Q^2 \rightarrow \infty)=2 f_\pi$.~\cite{Lepage:1980fj}

The results calculated with (\ref{eq:DressedFpiT2}) for $P_{q \bar q}=0.5$ are shown as dashed curves in Figs.  \ref{Q2PiTFF} and  \ref{PiTFF}.
One can see that the calculations with the dressed current are larger as compared with the results computed with the free current and the experimental data at low- and medium-$Q^2$
regions ($Q^2 <10$ GeV$^2$). The new results again disagree with \babar's data at large $Q^2$.

\section{Higher-Twist Components to the Transition Form Factor}
\label{sec:HigherTwist}

In a previous light-front QCD analysis of the pion TFF \cite{Brodsky:1980vj} it was argued that the valence Fock state $| q \bar q \rangle$ provides only half of the contribution to
the pion TFF at $Q^2=0$, while the other half comes from diagrams where the virtual photon couples inside the pion
(strong interactions occur between the two photon interactions). 
This leads to a surprisingly small value for the valence Fock state probability $P_{q \bar q}=0.25$.
More importantly, this raises the question on the role played by the higher Fock components of the pion LFWF,
\begin{equation}
\vert \pi \rangle =  \psi_2 \vert q \bar q \rangle + \psi_3 |q \bar q g \rangle +  \psi_4 | q \bar q q \bar q \rangle + \cdots,
\label{eq:Fockexp}
\end{equation}
in the calculations for the pion TFF.

The contributions to the transition form factor from these higher Fock states are suppressed,  compared with the valence Fock state, by the factor $1/(Q^2)^n$ 
for $n$ extra  $q \bar q$ pairs in the higher Fock state, since one needs to evaluate an off-diagonal matrix element between the real photon
and the multiquark Fock state.~\cite{Lepage:1980fj}
We note that in the case of the elastic form factor, the power suppression is $1/ (Q^2)^{2n}$ for $n$ extra $q \bar q$ pairs in the higher Fock state.
These higher Fock state contributions are negligible at high $Q^2$. 
On the other hand, it has  long been argued  that the higher Fock state
contributions are necessary to explain the experimental data at the medium $Q^2$ region for exclusive 
processes.~ \cite{CaoHM96,CaoCHM97}
The contributions from the twist-3 components of the two-parton pion distribution amplitude to the pion elastic form factors were evaluated 
in Ref. \cite{CaoDH99}.  The three-parton contributions to the pion elastic  form factor were studied in Ref. \cite{ChenL11}.
The contributions from diagrams where the virtual photon couples inside the pion to the pion transition form factor were estimated using light-front wavefunctions 
in Refs. \cite{HuangW07,WuH10}.
The higher twist (twist-4 and twist-6) contributions to the pion transition form factor~\cite{Gorsky87} were evaluated using the method of light-cone sum rules
in Refs.~\cite{SAgaevBOP11,BakulevMPS11},
 but opposite claims were made on whether the \babar\ data could be accommodated by including these higher twist contributions.

It is also not very clear how the higher Fock states contribute to decay processes, such as $\pi^0 \rightarrow \gamma \gamma$, \cite{AlkoferR92}
due to the long-distance nonperturbative nature of decay processes.
Second order radiative corrections to the triangle anomaly do not change the anomaly results as they contain one internal photon line and two vertices on the triangle loop.
Upon regulation no new anomaly contribution occurs.
In fact, the result is expected to be valid at all orders in perturbation theory.~\cite{Adler:1969er, Zee:1972zt}
It is thus generally argued that in the chiral limit of QCD ({\it i.e.}, $m_q \rightarrow 0$), one needs only the $q \bar q$ component to explain the anomaly, but as shown below,
the higher Fock state components
 can also contribute to the decay process $\pi^0 \to \gamma \gamma$ in the chiral limit.

As discussed in the last two sections, matching the AdS/QCD results computed with the free and dressed currents for the TFF at $Q^2=0$ with the anomaly result
requires a probability $P_{q \bar q}=0.5$. Thus it is important to investigate the contributions from the higher Fock states.
In AdS/QCD there are no dynamic gluons and confinement is realized via an effective instantaneous interaction
in light-front time, analogous to the instantaneous gluon exchange.~\cite{Brodsky:1997de}
The effective confining  potential also creates quark-antiquark pairs from the amplitude $q \to q \bar q q$.
Thus  in AdS/QCD higher Fock states can have any number of extra $q \bar q$ pairs. 
These higher Fock states lead to higher-twist contributions to the pion transition form factor.

\begin{figure}[h!]
\begin{center}
\includegraphics[width=12cm]{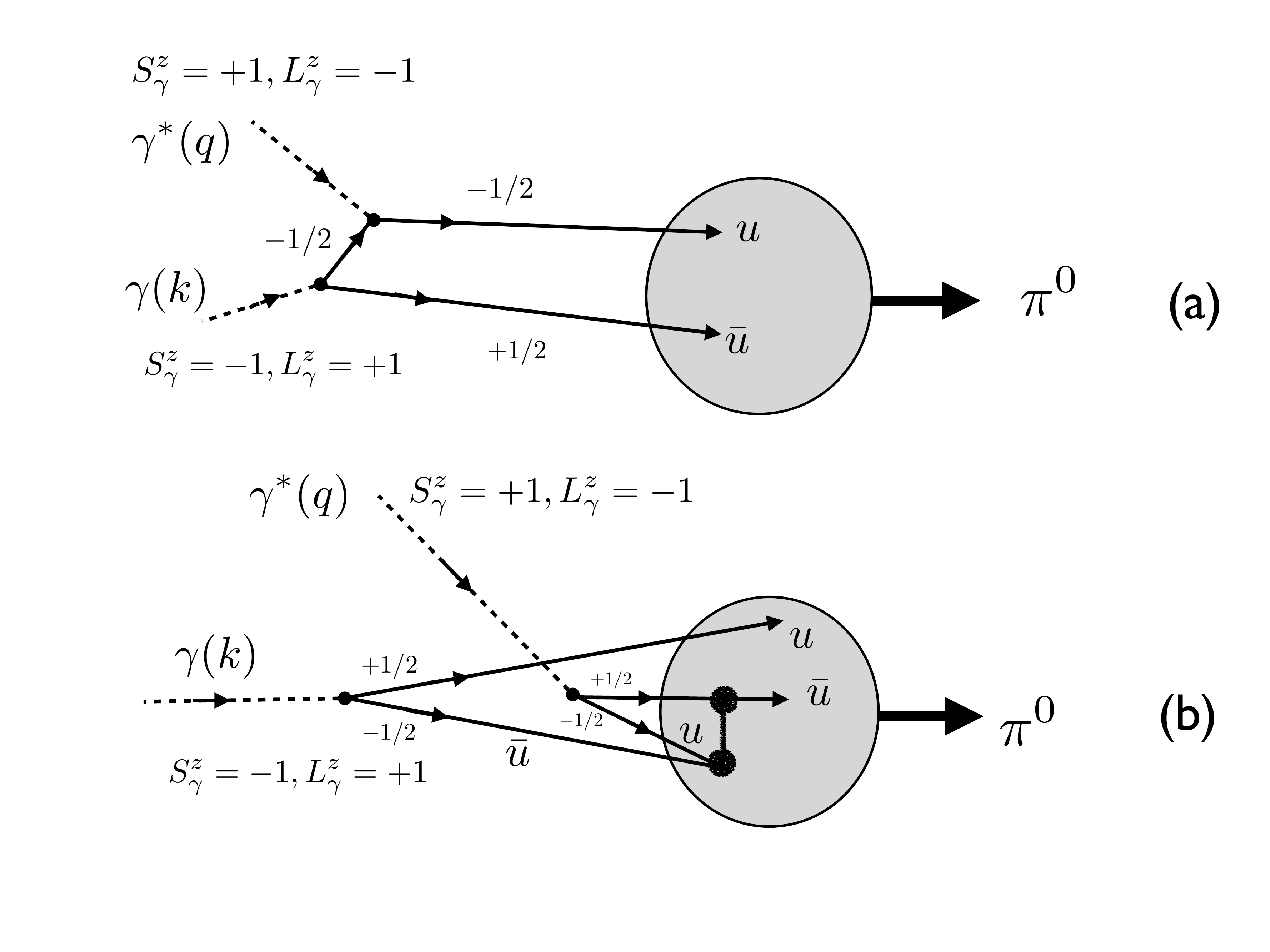}
\caption{Leading-twist contribution (a) and twist-four contribution (b) to the process $\gamma \gamma^* \to \pi^0$.}
\label{TFF}
\end{center}
\end{figure}

To illustrate this observation consider the two diagrams in Fig.~\ref{TFF}.
In the leading process,  Fig.~\ref{TFF} (a), where  both photons couple to the same quark, the valence
$|q \bar q \rangle$ state has  $J^z = S^z = L^z =0$,
\begin{equation} \label{eq:2q}
\vert q \bar q \rangle = \frac{1}{\sqrt 2} \Big( \Big\vert + \half, - \half \Big\rangle -  \Big\vert - \half, + \half \Big\rangle 
\Big).
\end{equation}
Equation~(\ref{eq:2q}) represents a $J^{PC} = 0^{-+}$ state with the quantum numbers of the 
conventional $\pi$ meson 
axial vector interpolating operator $\mathcal{O} = \bar \psi \gamma^+ \gamma^5 \psi$.

In the process involving the four-quark state $\vert q \bar q q \bar q\rangle$ of the pion, Fig.~\ref{TFF} (b), where each photon
couples directly to a $q \bar q$ pair, the four-quark state also satisfies $J^z = S^z = L^z =0$ and is represented by 
\begin{multline} \label{eq:4q}
\vert q \bar q q \bar q \rangle = \frac{1}{ 2} \Big( 
     \Big\vert + \half, - \half, + \half, - \half \Big\rangle 
+  \Big\vert + \half, - \half, - \half, + \half \Big\rangle \\
-   \Big\vert - \half, + \half, + \half, - \half \Big\rangle
-   \Big\vert - \half, + \half, - \half,  + \half \Big\rangle \Big).
\end{multline}
The four-quark state in Eq.~(\ref{eq:4q}) has also quantum numbers $J^{PC} = 0^{-+}$ corresponding to  the quantum numbers of the local interpolating operators
$\mathcal{O} = \bar \psi \gamma^+ \gamma^5 \psi \bar \psi \psi$ where 
the  scalar interpolating operator $\bar \psi \psi$ has quantum numbers $J^{PC} = 0^{++}$. 

We note that for the Compton scattering $\gamma H \to \gamma H$ process, similar higher-twist contributions, as illustrated in Fig.~\ref{TFF} (b),
are proportional to $\sum_{e_i \ne e_j} e_i e_j$ and are necessary to derive the low-energy amplitude for Compton scattering
which is proportional to the total charge squared $e_H^2 = (e_i + e_j)^2$ of the target. \cite{Brodsky:1968ea}

Both processes illustrated in Fig~(\ref{TFF}) make contributions to the two photon process $\gamma^* \gamma \to \pi^0$.
Time reversal invariance means that the four-quark state $\vert q \bar q q \bar q\rangle$ should also contribute to the decay process $\pi^0 \to \gamma \gamma$.
In a semiclassical model without dynamic gluons, Fig.~\ref{TFF} (b) represents the only higher twist term which contribute to the $\gamma^* \gamma \to \pi^0$ process.
The twist-four contribution vanishes at large $Q^2$ compared to the leading-twist  contribution, thus maintaining the asymptotic predictions while only modifying
the large distance behavior of the wave function.

To investigate the contributions from the higher Fock states in the pion LFWF, we write the twist-2 and twist-4 hadronic AdS components from (\ref{eq:Phitau})
\begin{eqnarray}  \label{eq:t2}
\Phi_\pi^{\tau = 2}(z) &=& \frac{\sqrt 2 \kappa z^2 }{\sqrt{1 + \alpha^2}}  e^{- \kappa^2 z^2/2}, \\ \label{eq:t4}
\Phi_\pi^{\tau = 4}(z) &=&  \frac{\alpha \kappa^3 z^4 }{\sqrt{1 + \alpha^2}}  e^{- \kappa^2 z^2/2},
\end{eqnarray}
with normalization
\begin{equation}
\int_0^\infty \frac{dz}{z^3} \left[\vert \Phi_\pi^{\tau = 2}(z) \vert^2 + \vert \Phi_\pi^{\tau = 4}(z) \vert^2 \right] = 1,
\label{eq:normT24}
\end{equation}
and probabilities $P_{q \bar q}=1/(1+ \vert \alpha \vert^2)$ and $P_{q \bar q q \bar q}=\alpha^2/(1+ \vert \alpha \vert^2)$.
The pion decay constant 
follows from the short-distance asymptotic behavior of the leading contribution
and is given by
\begin{equation}  \label{eq:fpialpha}
f_\pi = \frac{1}{\sqrt{1 + \alpha^2}} \frac{\kappa}{ \sqrt 2 \pi}.
\end{equation}

Using (\ref{eq:t2}) and (\ref{eq:t4}) together with (\ref{eq:Vx}) in equation (\ref{eq:TFFAdS2}) we find the total contribution from twist-2 and twist-4 components for the 
dressed current,
\begin{equation}  
F_{\pi \gamma}(Q^2) =  \frac{1 }{\pi^2 f_\pi}  \frac{1}{(1 + \alpha^2)^{3/2}} \int_0^1  \frac{d x }{(1+x)^2} 
x^{Q^2 / [8 \pi^2 f_\pi^2 (1+\alpha^2)]}
\left[1 + \frac{4 \alpha}{\sqrt 2} \frac{1-x}{1+x} \right].
\label{eq:DressedFpiT2+4}
\end{equation}
The transition form factor at $Q^2=0$ is given by
\begin{equation}
F_{\pi \gamma}(0)=\frac{1 }{2 \pi^2 f_\pi}  \frac{1+\sqrt{2} \alpha }{(1 + \alpha^2)^{3/2}}.
\label{eq:FpiTwist2+4Q0}
\end{equation}
The Brodsky-Lepage asymptotic prediction for the pion TFF can be recovered from Eq.~(\ref{eq:DressedFpiT2+4}) by noticing that 
the second term  
vanishes at $Q^2 \rightarrow \infty$ and the similarity between Eq.~(\ref{eq:DressedFpiT2Asy})
and the first term in Eq.~(\ref{eq:DressedFpiT2+4}).

Imposing the anomaly result (\ref{eq:JackiwAnomaly}) on (\ref{eq:FpiTwist2+4Q0}) we find two possible real solutions for 
$\alpha$: $\alpha_1=-0.304$ and $\alpha_2=1.568$.~\footnote{If we impose the condition that the twist 4 contribution at $Q^2=0$ is exactly
half the value of the twist-2 contribution one
  obtains $\alpha = - \frac{1}{2 \sqrt 2}$, which is very close to the value of $\alpha$ which follows by imposing the
  triangle anomaly constraint. In this case the pion TFF has a very simple form
  $F_{\pi \gamma}(Q^2) = \frac{8 }{3 \pi \kappa} \int_0^1  \frac{d x}  {(1+x)^3} \, x^{Q^2 / 4 \kappa^2 + 1}$.}
The larger value $\alpha_2=1.568$ yields $P_{q \bar q}=0.29$, $P_{q \bar q q \bar q}=0.71$, and $\kappa=1.43$ GeV. The  resulting value of $\kappa$ is about 4 times larger than the value
obtained from the AdS/QCD analysis of the hadron spectrum and the pion elastic form factor,~\cite{deTeramond:2010ez} and thereby should be discarded.
The other solution $\alpha_1=-0.304$ gives $P_{q \bar q}=0.915$, $P_{q \bar q q \bar q}=0.085$, and $\kappa=0.432$ GeV -- results that
are similar to that found from an analysis of the space and timelike behavior of the pion form factor using LF holographic methods,
including higher Fock components in the pion wave function.~\cite{deTeramond:2010ez}
Semiclassical holographic methods, where dynamical gluons are not presented, are thus compatible with a large probability for the valence state of the order of 90\%.
On the other hand, QCD analyses including multiple gluons on the pion wave function favor a small probability (25\%) for the valence 
state.~\cite{Brodsky:1980vj} Both cases (and examples in between) are examined in Ref.~\cite{Brodsky:2011xx}.

The results for the transition form factor are shown as solid curves in Figs.  \ref{Q2PiTFF} and  \ref{PiTFF}.
The agreements with the experimental data at low- and medium-$Q^2$ regions ($Q^2 <10$ GeV$^2$) are greatly improved compared with the results obtained with only
twist-two component computed with the dressed current. 
However, the rapid growth of the pion-photon transition form factor exhibited by the \babar\  data at high $Q^2$ still cannot be reproduced.
So we arrive at a similar conclusion as we did in a QCD analysis of the pion TFF in Ref.~\cite{Brodsky:2011xx}:
it is difficult to explain the rapid growth of the form factor exhibited by the \babar\  data at high $Q^2$ within the current framework of QCD.

\section{Transition Form Factors for the $\eta$ and $\eta^\prime$ Mesons}
\label{sec:eta}

The  $\eta$ and $\eta^\prime$ mesons result from the mixing of
the neutral states $\eta_8$ and $\eta_1$ of the SU(3)$_F$ quark model. The transition form factors for the latter have the same expression as  the pion transition form factor, 
except an overall multiplying factor $c_P=1, \, \frac{1}{\sqrt{3}}$, and $\frac{2\sqrt{2}}{\sqrt{3}}$ for the $\pi^0$, $\eta_8$ and $\eta_1$, respectively.
By multiplying  Eqs. (\ref{eq:TFFAdSQCD}), (\ref{eq:DressedFpiT2}) and (\ref{eq:DressedFpiT2+4})
by the appropriate factor  $c_P$, one obtains the corresponding expressions for the transition form factors for the $\eta_8$ and $\eta_1$.

\begin{figure}[htbp]
\begin{center}
\includegraphics[width=9cm]{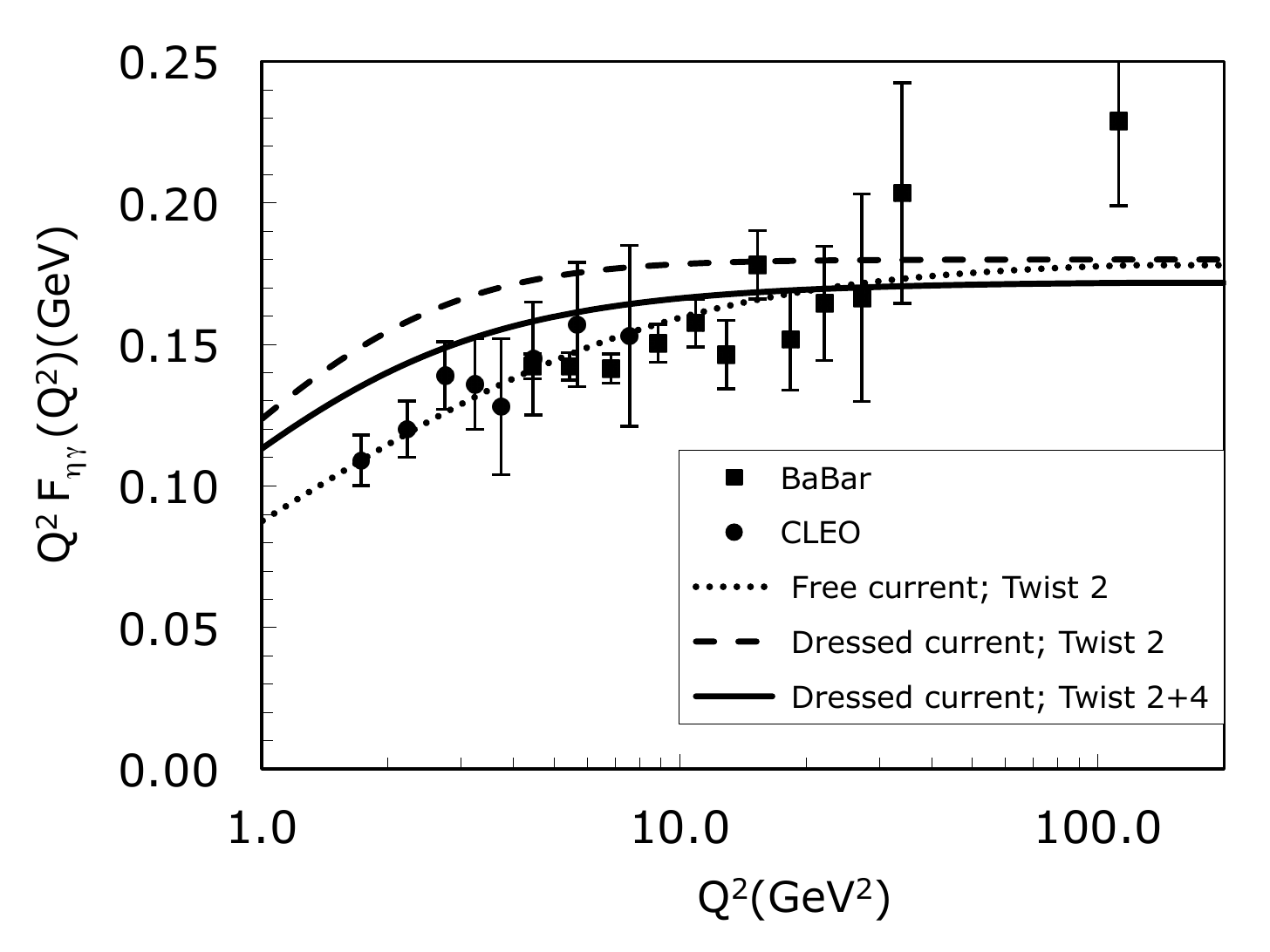}
\caption{The $\gamma \gamma^* \rightarrow \eta$ transition form factor shown as $Q^2 F_{\eta \gamma}(Q^2)$ as a function of $Q^2 = -q^2$.  The dotted curve is the asymptotic result.
The dashed and solid curves include the effects of using a confined EM current for twist-2 and twist-2 plus twist-4, respectively.
The data are from \cite{Aubert:2009mc, CELLO,  CLEO}. }\label{fig:EtaTFF_Q2Feta}
\end{center}
\end{figure}

\begin{figure}[htbp]
\begin{center}
\includegraphics[width=9cm]{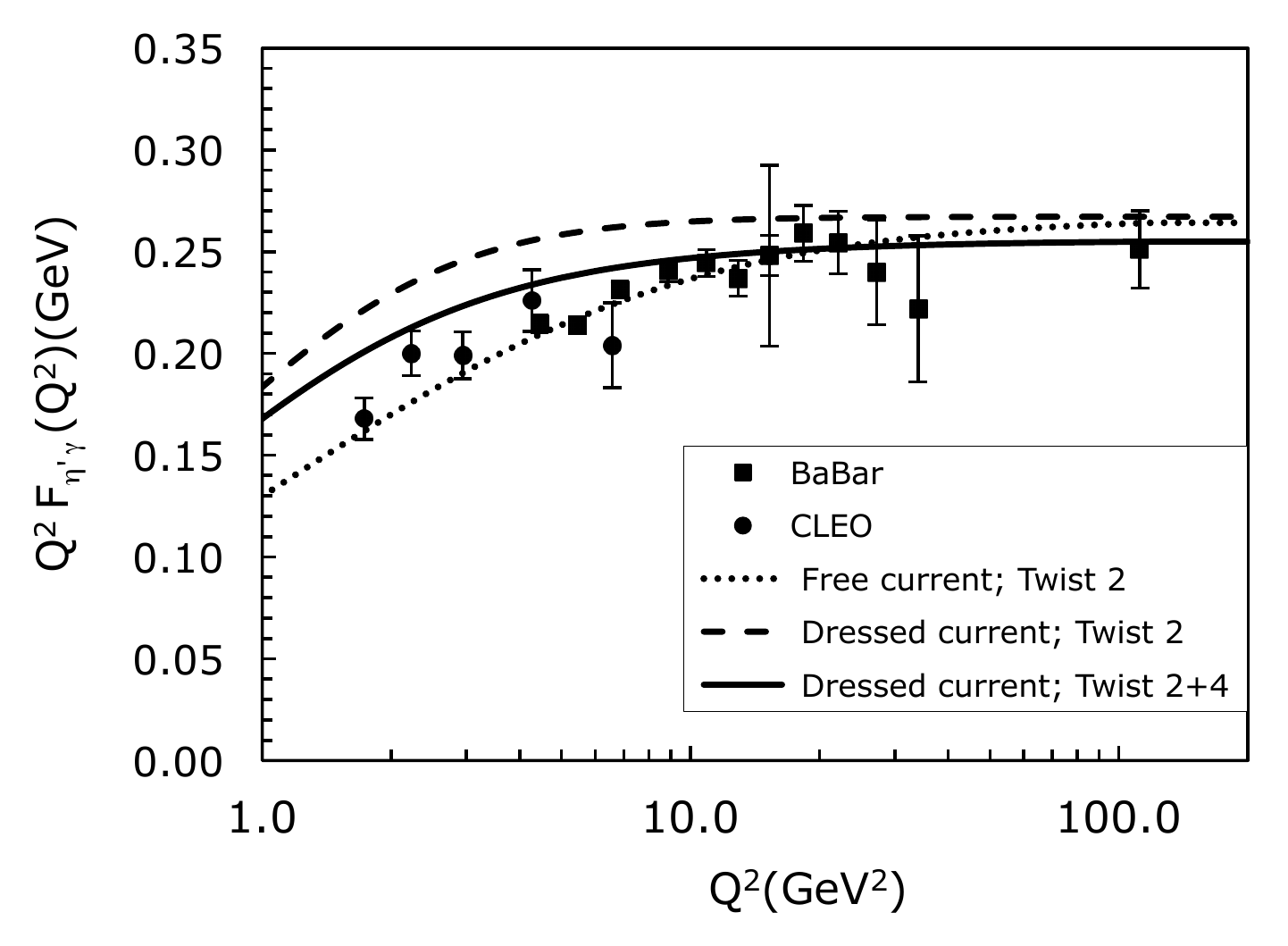}
\caption{Same as Fig. \ref{fig:EtaTFF_Q2Feta} for the $\gamma \gamma^* \rightarrow \eta^\prime$ transition form factor
shown as   $Q^2 F_{\eta^\prime \gamma}(Q^2)$.}
\label{fig:EtaPTFF_Q2FetaP}
\end{center}
\end{figure}

\begin{figure}[htbp]
\begin{center}
\includegraphics[width=9cm]{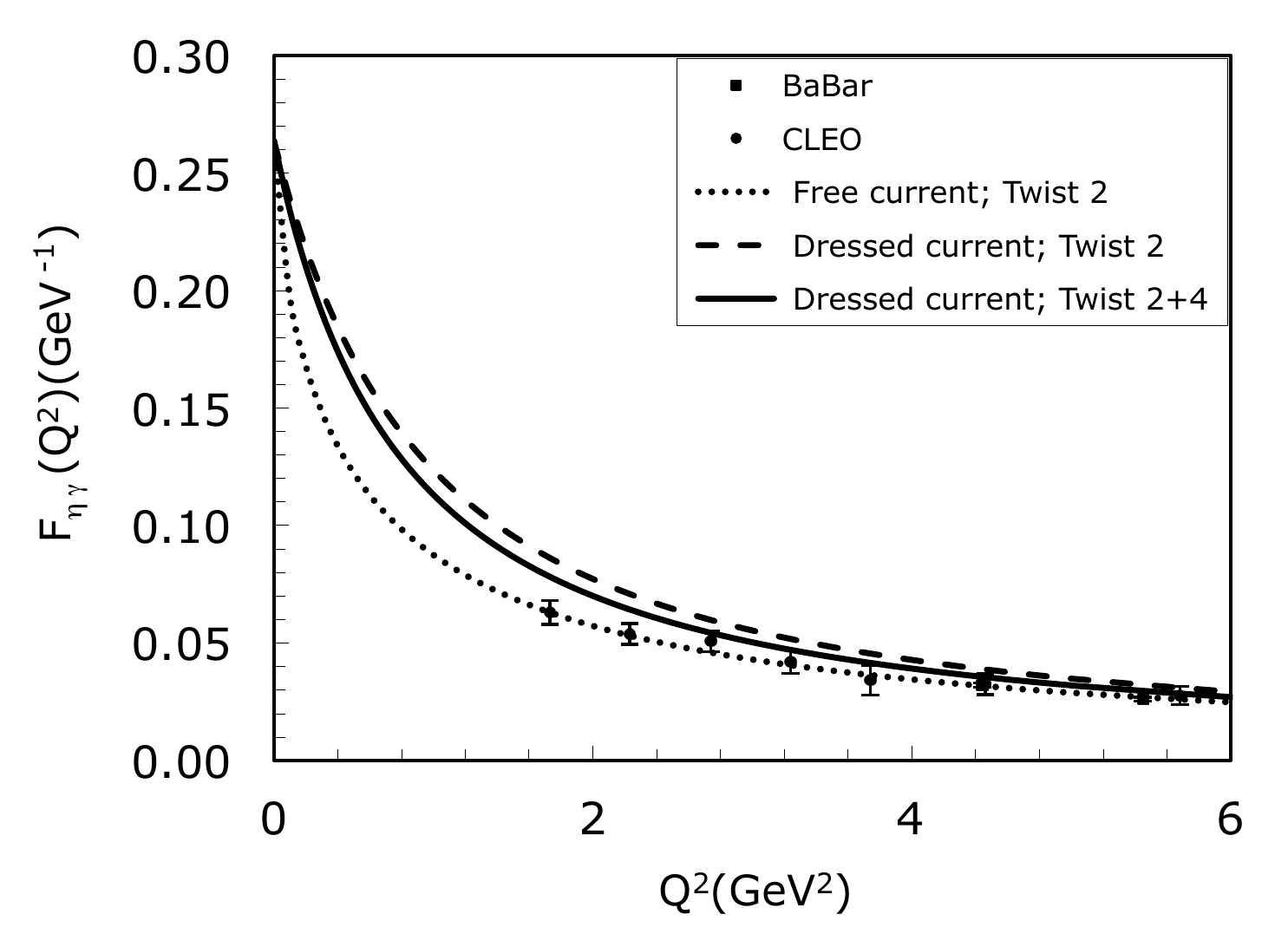}
\caption{Same as Fig. \ref{fig:EtaTFF_Q2Feta} for the $\gamma \gamma^* \rightarrow \eta$ transition form factor
shown as   $F_{\eta \gamma}(Q^2)$.}
\label{fig:EtaTFF_Feta}
\end{center}
\end{figure}

\begin{figure}[htbp]
\begin{center}
\includegraphics[width=9cm]{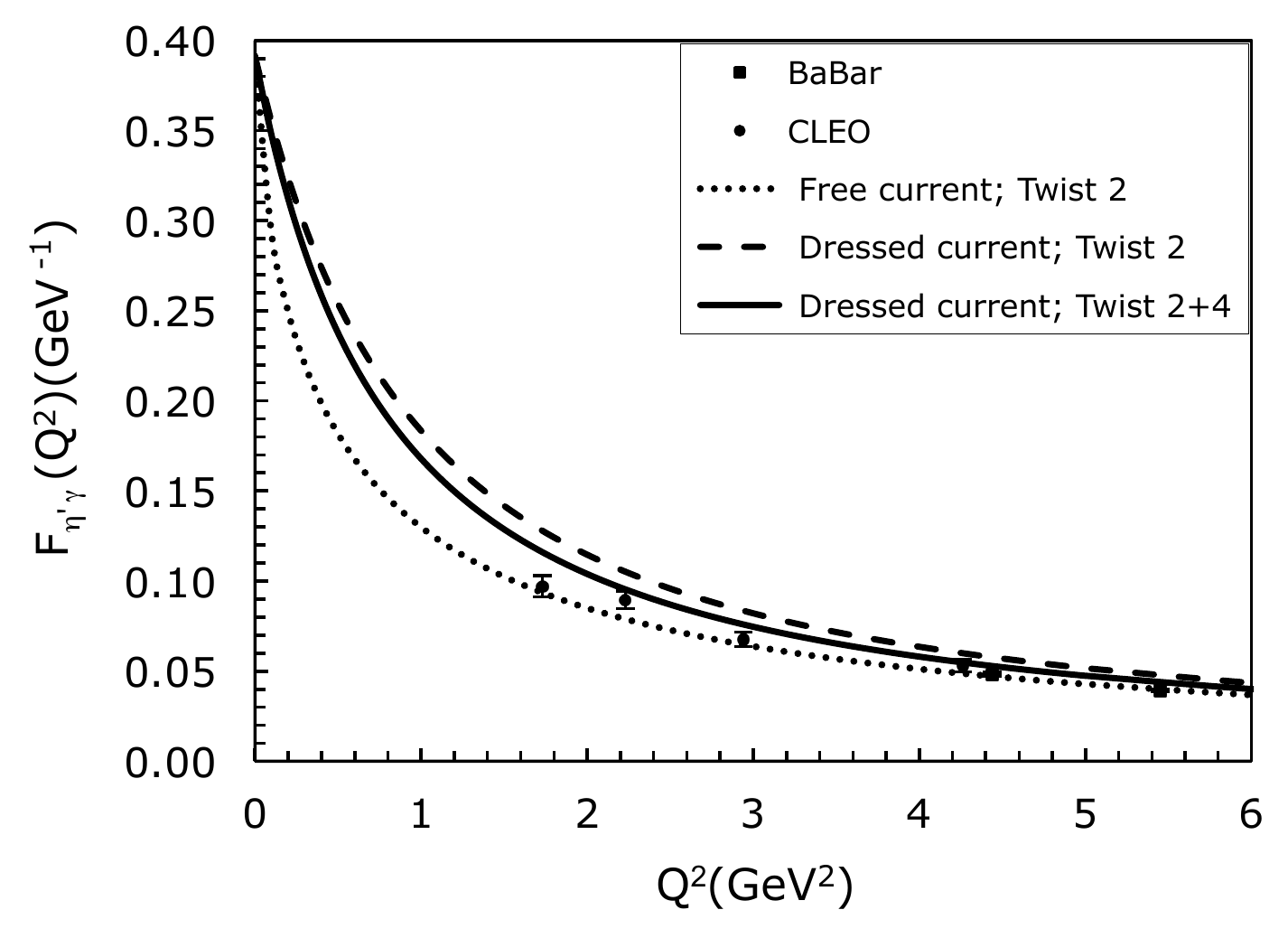}
\caption{Same as Fig. \ref{fig:EtaTFF_Q2Feta} for the $\gamma \gamma^* \rightarrow \eta^\prime$ transition form factor
shown as   $F_{\eta^\prime \gamma}(Q^2)$.}
\label{fig:EtaPTFF_FetaP}
\end{center}
\end{figure}

The transition form factors for the physical states  $\eta$ and $\eta^\prime$ are a superposition of the transition form factors for the  $\eta_8$ and $\eta_1$
\begin{eqnarray}
\left(
	\begin{array}{c}
	F_{\eta \gamma} \\
	F_{\eta^\prime \gamma}
	\end{array}
\right)
=\left(
	\begin{array}{cc}
	{\rm cos} \, \theta & -{\rm sin} \, \theta \\
	{\rm sin} \, \theta & {\rm cos} \, \theta
	\end{array}
\right)
\left(
	\begin{array}{c}
	F_{\eta_8 \gamma} \\
	F_{\eta_1 \gamma}
	\end{array}
\right),
\end{eqnarray}
where $\theta$ is the mixing angle for which we adopt $\theta=-14.5^o\pm 2^o$.~\cite{Cao:1999fs}
The results for the $\eta$ and $\eta^\prime$ transitions form factors are shown in
Figs. \ref{fig:EtaTFF_Q2Feta} and \ref{fig:EtaPTFF_Q2FetaP} for $Q^2 F_{M\gamma}(Q^2)$, and Figs. \ref{fig:EtaTFF_Feta} and \ref{fig:EtaPTFF_FetaP} for $F_{M\gamma}(Q^2)$.
The calculations agree very well with available experimental data over a large range of $Q^2$.
We note that other mixing schemes were proposed in studying the mixing behavior of the decay constants and states of the $\eta$ and $\eta^\prime$
mesons. \cite{Leutwyler98,FeldmannK98,FeldmannKS98}
Since the transition form factors are the primary interest in this study it is appropriate to use the conventional single-angle mixing scheme
for the states. Furthermore, the predictions for the $\eta$ and $\eta^\prime$ transition form factors remain largely unchanged if other mixing schemes are used in the calculation.

\section{Conclusions}
\label{sec:Conclusions}

The light-front holographic approach  provides a direct mapping between an effective gravity theory defined in a fifth-dimensional warped space-time
and a corresponding  semiclassical approximation to strongly coupled QCD quantized on the light-front.
In addition to predictions for hadron spectroscopy,  important outputs are the elastic form factors of hadrons and constraints on their light-front bound-state wave functions.
The soft-wall holographic  model is particularly successful.

We have studied the photon-to-meson transition form factors $F_{M \gamma}(Q^2)$ for $\gamma^* \gamma \to M$ using light-front holographic methods. 
The Chern-Simons action, which is a natural form in five-dimensional AdS space, is required to describe the anomalous coupling of mesons to photons using holographic methods and leads directly to an expression for the photon-to-pion transition form factor for a class of confining models. 
Remarkably, the pion transition form factor given by Eq. (\ref{eq:TFFAdSQCD}) derived from the CS action is identical to the leading order QCD result where 
the distribution amplitude has the asymptotic form $\phi(x) \propto x(1-x).$

The Chern-Simons form is  local in AdS space and is thus somewhat limited in its predictability. It  only retains the $q \bar q$ component of the pion wave function,
and further, it projects out only the asymptotic form of the meson distribution amplitude $\phi(x) \propto x(1-x)$. In contrast, the 
holographic  light-front mapping of electromagnetic and gravitational form factors  gives the full form of the distribution amplitude $\phi(x) \propto \sqrt{x(1-x)}$
for arbitrary values of $Q^2$.
This apparently contradictory result was first found in Ref.~\cite{Grigoryan:2008up} in
a hard-wall AdS extended model.  This contradiction indicates that the local interaction from the CS action can only represent the pointlike asymptotic form.
The asymptotic result coincides with the CS amplitude which is only sensitive  to short-distance physics.
 If the QCD evolution for the distribution amplitude $\phi \propto \sqrt{x(1-x)}$ is included  the asymptotic DA is recovered at very large $Q$.~\cite{Brodsky:2011xx}

It is found that in order to describe simultaneously the decay process $\pi^0 \rightarrow \gamma \gamma$ and the pion TFF at the asymptotic limit a probability for
the $q \bar q$ component of the pion wave function $P_{q \bar q}=0.5$ is required for the calculations with the free and dressed AdS currents.

We have argued that the contributions from the higher Fock components in the pion light-front wave function also 
need to be included in the analysis of exclusive processes.
In fact, just as in 1+1 QCD, the confining interaction of the LF Hamiltonian  in light-front holography
 leads to Fock states with any number of extra $q \bar q$ pairs.
These contributions lead to higher-twist contributions to the hadron form factor. 
We have shown how the effect of the higher Fock states in form factors can be obtained by analyzing the hadron matrix elements of the confined dressed electromagnetic
Heisenberg current from the gauge/gravity duality.
The probability for the four-quark states obtained in this work, $P_{q \bar q q \bar q} = 0.085$ is similar to that found from an analysis
of the spacelike and timelike behavior of the pion form factor using LF holographic methods, including higher Fock components in the pion wave function.~\cite{deTeramond:2010ez}

The results obtained for the $\eta$- and $\eta^\prime$-photon transition form factors
are consistent with all currently available experimental data.  
However, the rapid growth of the pion-photon transition form factor exhibited by the \babar\  data at high $Q^2$ is not compatible with
the models discussed in this article, and in fact is very difficult to explain within the current framework of QCD.

\acknowledgements

We thank C. D. Roberts for helpful comments. This research was supported by the Department of Energy Contract
No. DE--AC02--76SF00515.

\appendix

\section{Light-Front Wave functions From  Holographic Mapping \label{appendix}}
\label{sec:LFWFs}

For a two-parton bound state, light-front holographic mapping relates the light-front wave function  
$\psi(x, \zeta, \varphi)$ in physical space-time  with its dual   field $\Phi(z)$ in AdS space. The precise relation is given by (\ref{psiphi}) 
\begin{equation}    \label{psiPhi}
\psi(x,\zeta, \varphi) = e^{i M \varphi} X(x) \frac{\phi(\zeta)}{\sqrt{2 \pi \zeta}},
\end{equation}
where  we have factored out
the angular dependence $\varphi$ and the longitudinal, $X(x)$, and transverse mode~\cite{Brodsky:2006uqa, Brodsky:2007hb, Brodsky:2008pf, deTeramond:2008ht}  
\begin{equation} \label{phiPhi}
\phi(\zeta) = \zeta^{-3/2} \Phi(\zeta).
\end{equation}
The holographic variable $z$ is related to the light-front  invariant variable $\zeta$ which represents
the transverse separation of the quarks within the pion
\begin{equation} \label{zetaz}
z \to \zeta = \sqrt{x(1-x)} \vert \mbf b_\perp \vert.
\end{equation}
The LF variable $x$ is the longitudinal light-cone momentum fraction $x = k^+/P^+$ and $\mbf{b}_\perp$ is the impact separation and Fourier conjugate to $\mbf{k}_\perp$, the relative transverse momentum coordinate.

The LFWF is normalized according to
\begin{equation}
\langle \psi_{q \bar q/\pi} \vert \psi_{q \bar q/\pi} \rangle = P_{q \bar q},
\end{equation}
where $P_{q \bar q}$ is the probability of finding the $q \bar q$ component in the pion light-front wave function.
We choose the normalization of the LF mode $\phi(\zeta) = \langle \zeta\vert \psi\rangle$ as 
\begin{equation}
\langle \phi \vert \phi \rangle =\int d\zeta \, \vert \langle  \zeta \vert \phi \rangle \vert^2 = P_{q \bar q},
\end{equation}
and thus the longitudinal mode is normalized as
\begin{equation}
\int_0^1 \frac{X^2(x)}{x(1-x)}=1.
\end{equation}

As we have shown in Sec. \ref{sec:MFF}   the factorization (\ref{psiPhi}) is required to map the elastic electromagnetic
 form factors for arbitrary values of the transverse momentum $Q$ with the result $X(x) = \sqrt{x(1-x)}$~\cite{Brodsky:2006uqa, Brodsky:2007hb} for the longitudinal mode. Identical results follow from the mapping to
 the gravitational form factor.~\cite{Brodsky:2008pf} The longitudinal mode $X(x)$ cannot be determined from the mapping of the Hamiltonian equation for bound states as it decouples in the ultra relativistic limit  $m_q \to 0$.~\cite{deTeramond:2008ht} 

For a harmonic confining potential $U(z) \sim \kappa^4 z^2$ we have from  (\ref{eq:Phitau})  
\begin{equation}   \label{eq:PhipiApp}
\Phi_{q \bar q/ \pi} (z) =   \sqrt{2 P_{q \bar q}} \, \kappa \, z^2 e^{- \kappa^2 z^2/2},
\end{equation} 
for a twist $\tau = 2$ mode propagating in AdS space. From Eqs.  (\ref{psiPhi}),  (\ref{phiPhi}) and
 (\ref{zetaz}) we find the LFWF   ($L^z = M = 0$)
\begin{equation} 
\psi_{q \bar q/\pi}(x, \mbf{b}_\perp)  
= \frac{\kappa}{\sqrt{\pi}}   \sqrt{P_{q \bar q}} \sqrt{x(1-x)}~e^{-\half \kappa^2 x(1-x) \mbf{b}_\perp^2}.
\label{eq:psiqqbar}
\end{equation}
in physical space time.

The pion distribution amplitude in the light-front formalism~\cite{Lepage:1980fj} is the integral of the 
valence $q \bar q$ light-front wave function
\begin{equation}
\phi(x)=\int \frac{ d^2 \mbf{k}_\perp}{16 \pi^3} \psi_{q \bar q/ \pi}(x, \mbf{k}_\perp),
\label{eq:DALCApp}
\end{equation}
and satisfies the normalization condition which follows from the decay process $\pi \to \mu \nu$ ($N_C = 3$)
\begin{equation}
\int_0^1 {\rm d}x \, \phi(x)  = \frac{f_\pi}{2\sqrt{3}},
\label{Eq:DAnormalization}
\end{equation}
where  $f_\pi = 92.4$ MeV is the pion decay constant. From (\ref{eq:psiqqbar}) we find
the distribution amplitude 
\begin{equation}
\phi(x) = \frac{4}{\sqrt 3 \pi} \sqrt{x(1-x)},
\end{equation}
and the pion decay constant
\begin{equation}
f_\pi = \sqrt{P_{q \bar q}} \frac{\sqrt 3}{8} \kappa.
\end{equation}

As discussed in the paper, the CS mapping gives the asymptotic distribution amplitude since the CS maps a pointlike pion. The corresponding longitudinal mode in the LFWF
is $X(x) = \sqrt{6} \, x (1-x)$ and thus the LFWF
\begin{equation} 
\psi_{q \bar q/\pi}(x, \mbf{b}_\perp)  
= \frac{\kappa}{\sqrt{\pi}}   \sqrt{P_{q \bar q}} \sqrt{6}\, x(1-x)~e^{-\half \kappa^2 x(1-x) \mbf{b}_\perp^2}.
\label{eq:psiqqbarasy}
\end{equation}
The pion decay constant in this case is 
\begin{equation}
f_\pi = \sqrt{P_{q \bar q} } \, \frac{\kappa}{\sqrt{2} \pi},
\end{equation}
consistent with (\ref{eq:fpiPhi}).

The evolution of the pion distribution amplitude in $\log Q^2$ is governed by the Efremov-Radyushkin-Brodsky-Lepage (ERBL) evolution
equation~\cite{Lepage:1980fj, Efremov:1979qk}. It can be expressed in terms of 
the anomalous dimensions of  the Gegenbauer polynomial projection of the DA.
If we normalize the full LFWF of the pion by $\langle \psi \vert \psi \rangle = 1$, we can compute the probability  to find the pion in a given component of
 a Gegenbauer polynomial expansion
$X(x) = x(1-x) \sum_n \alpha_n C_n^{(3/2)}(2x - 1)$.
We find
\begin{equation} \label{Pn}
P_n = \frac{(n + 2)(n+1)}{4(2 n + 3)}  \alpha_n^2,
\end{equation}
where $\sum_n P_n = 1$. For the AdS solution $X(x) = \sqrt{x(1-x)}$ the asymptotic component $\alpha_0 = 3 \pi / 4$ and the probability to find the pion in its asymptotic state  is 
$P_0 = 3 \pi^2/32 \simeq 92.5 \, \%$, not too far from the asymptotic result. Notice that
 $P_n$ in (\ref{Pn}) are the probabilities related to the Gegenbauer projection of the valence state of the pion.  
They are not related to the probabilities discussed in the section below, which are the probabilities of  higher particle number Fock states in the pion.

The asymptotic form has zero anomalous dimension. The distribution amplitude  $\phi(x) \propto \sqrt{x(1-x)}$ derived from LF holographic methods is sensitive 
to soft physics  $1-x \sim \kappa/Q^2$, and has Gegenbauer polynomial components with nonzero anomalous dimensions which are driven to zero for large values of $Q^2$.
Expanding the distribution amplitude at any finite scale as $x(1-x)$ times Gegenbauer polynomials, only its projection on the lowest Gegenbauer
polynomial with zero anomalous moment survives at large $Q^2$.

\subsection{Effective Light-Front Wave Function From Holographic Mapping of a Confined Electromagnetic Current }

It is also possible to find a precise mapping of a confined EM current propagating in a warped AdS space to the light-front QCD Drell-Yan-West expression for the form factor. 
In this case the resulting LFWF  incorporates non valence higher Fock states generated by the ``dressed'' confined current. For the soft-wall model this mapping
can be done analytically.   

The form factor in light-front  QCD can be expressed in terms of an effective single-particle density~\cite{Soper:1976jc}
\begin{equation} 
F(Q^2) =  \int_0^1 dx \, \rho(x,Q),
\end{equation}
where
\begin{equation} \label{rhoQCD}
\rho(x, Q) = 2 \pi \int_0^\infty \!  b \,  db \, J_0(b Q (1-x)) \vert \psi(x,b)\vert^2,
\end{equation}
for a two-parton state ($b = \vert \mbf{b}_\perp \vert$).

 We can also compute an effective density on the gravity side corresponding to a twist $\tau$ hadronic mode $\Phi_\tau$ in a modified AdS space.
 For the soft-wall model the expression is~\cite{Brodsky:2007hb}
 \begin{equation}  \label{rhoAdS}
\rho(x,Q) = (\tau \!-\!1) \, (1 - x)^{\tau-2} \, x^{\frac{Q^2}{4 \kappa^2}} .
\end{equation}
To compare (\ref{rhoAdS}) with the QCD expression (\ref{rhoQCD}) for twist-2 we use the integral
\begin{equation}
\int_0^\infty \! u \, du  \, J_0(\alpha u) \,e^{- \beta u^2} = \frac{1}{2 \beta} \, e^{-\alpha^2/4\beta},
\end{equation}
and the relation $x^\gamma  = e^{\gamma \ln(x)}$. We find the effective two-parton  LFWF
\begin{equation} \label{ELFWF}
\psi(x, \mbf{b}_\perp) = \kappa \frac{ (1-x)}{\sqrt{\pi \ln(\frac{1}{x})}} \,
e^{- \half \kappa^2 \mbf{b}_\perp^2  (1-x)^2 / \ln(\frac{1}{x})},
\end{equation}
in impact space. The momentum space expression follows from the Fourier transform of  (\ref{ELFWF})
and it is given by  
\begin{equation} 
\psi(x, \mbf{k}_\perp) = 4 \pi \, \frac{ \sqrt{\ln\left(\frac{1}{x}\right)}}{\kappa (1-x)} \,
x^{\mbf{k}_\perp^2/2 \kappa^2 (1-x)^2}. 
\end{equation}
The effective LFWF  encodes  nonperturbative dynamical aspects that cannot be learned from a term-by-term holographic mapping, unless one adds an infinite number of terms.
Furthermore, it has the right analytical properties to reproduce the bound-state vector meson pole in the timelike EM form factor. Unlike the ``true'' valence LFWF,
the effective LFWF, which represents a sum of an infinite number of Fock components, is not symmetric in the longitudinal variables $x$ and $1-x$
for the active and spectator quarks,  respectively.

As we have discussed in Secs.  \ref{TFFSWM} and \ref{sec:TFFDressedCurrent} for the free and dressed currents, respectively, a simple model
with only a twist-2 valence pion state  requires a 50 \% probability, $P_{q \bar q} = \frac{1}{2}$,  to reproduce the decay process $\pi^0 \to \gamma \gamma$.  We recall that 
for the soft-wall model the EM form factor is given by (\ref{F}), and thus  for $\tau =2$ its asymptotic normalization is given by 
\begin{equation}  \label{FasyP}
Q^2 F(Q^2 \to \infty) = P_{q \bar q} M_\rho^2.
\end{equation}

One of the unsolved difficulties of the holographic approach to QCD is that the vector mesons masses obtained from the spin-1 equation of motion does not match the poles
of the dressed current when computing a form factor.
The discrepancy is an overall  factor of $\sqrt 2$.~\footnote{This discrepancy is also present in the gap scale if one computes the spectrum and  form factors without recourse to
 holographic methods, for example using the semiclassical approximation of Ref. \cite{deTeramond:2008ht}. In this case a discrepancy of a factor $\sqrt 2$
 is also found between the spectrum and the computation of spacelike form factors.}
Light front holography provides a precise relation of the fifth-dimensional mass $\mu$ with the total and orbital angular momentum of a hadron in the  transverse LF plane   
$(\mu R)^2 = - (2 - J)^2 + L^2$, $L = \vert L^z\vert$.~\cite{deTeramond:2008ht} Thus the $\rho$ meson mass  computed from the AdS wave equations for
a conserved  current $\mu R = 0$, corresponds  to a $J = L = 1$ twist-3 state.  In fact, the twist-3 computation of the spacelike form factor,  involves the current $J^+$,
and the poles do not correspond to the physical poles of the twist-2 transverse current $\mbf{J}_\perp$ present in the annihilation channel,
namely the $J = 1, L = 0$ state.~\cite{deTeramond:2011qp}  

If we define the physical vector meson mass by the relation $\overline{M}_\rho = P_{q \bar q}  M_\rho= M_\rho/\sqrt{2}$ the asymptotic result
(\ref{FasyP}) becomes
\begin{equation}  \label{Fasy}
Q^2  F(Q^2 \to \infty) = \overline{M}_\rho^2.
\end{equation}
We can thus define a form factor $\overline{F}(Q^2)$ shifting the poles in (\ref{F}) to their physical locations but keeping the same analytical structure.  Thus
for $\tau = 2$
\begin{equation}
 \overline{F}(Q^2) = \frac{1}{1+ \frac{Q^2}{\overline{M}_\rho^2}},
 \end{equation}
 which satisfies the asymptotic normalization (\ref{Fasy}) and charge normalization at $Q=0$,
 $F(0) = 1$.   For arbitrary twist $\tau$ the expression is
 \begin{equation} \label{Fbar}   
\overline{F}_\tau(Q^2) =  \frac{\overline{P}_\tau}{{\Big(1 + \frac{Q^2}{\overline{M}^2_\rho} \Big) }
 \Big(1 + \frac{Q^2}{\overline{M}^2_{\rho'}}  \Big)  \cdots 
       \Big(1  + \frac{Q^2}{\overline{M}^2_{\rho^{\tau-2}}} \Big)} .
\end{equation}

It is important to notice that the values of the probabilities $\bar{P}$ corresponding to the physical vector masses
$\overline{M}$ are markedly different from the probabilities  $P$ obtained from the formulas with the unphysical masses $M$. 
For example for a pion  $\overline{P}_{q \bar q} \simeq 90 \, \%$ and $\overline{P}_{q \bar q q \bar q} \simeq 10 \, \%$.~\cite{deTeramond:2010ez}
 When the vector meson masses are shifted to their physical values the agreement of the predictions with observed data
 is very good.~\cite{deTeramond:2010ez, deTeramond:2011qp}  Although the arguments presented above are not rigorous,
  they can help explain why a systematic difference of a factor $\sqrt{2}$ in the gap scale is found when comparing predictions with the spectrum or form factor data.

\end{document}